\documentclass[prb,amsmath,twocolumn,superscriptaddress,showpacs,nofootinbib]{revtex4-1}
\usepackage{epsfig}
\usepackage{amssymb}
\usepackage{amsmath}
\usepackage{amsfonts}
\usepackage[T1]{fontenc}
\usepackage{bm}
\usepackage{graphicx}
\usepackage{color}
\usepackage{verbatim}

\begin{document}
\title{Feynman Path Integrals Over Entangled States}

\author{A.~G. Green}
\affiliation{London Centre for Nanotechnology, University College London, Gordon St., London, WC1H 0AH, United Kingdom}

\author{C.~A. Hooley}
\affiliation{SUPA, School of Physics and Astronomy, University of St Andrews, North Haugh, St Andrews, Fife, KY16 9SS}

\author{J. Keeling}
\affiliation{SUPA, School of Physics and Astronomy, University of St Andrews, North Haugh, St Andrews, Fife, KY16 9SS}

\author{S.~H. Simon}
\affiliation{The Rudolf Peierls Centre for Theoretical Physics, 1 Keble Rd., Oxford, OX1 3NP}

\date{\today}
\begin{abstract}
The saddle points of a conventional Feynman path integral are not entangled, since they comprise a sequence of classical field configurations. We combine insights from field theory and tensor networks by constructing a Feynman path integral over a sequence of matrix product states. The paths that dominate this path integral include some degree of entanglement. This new feature allows several insights and applications: i. A Ginzburg-Landau description of deconfined phase transitions. ii. The emergence of new classical collective variables in states that are not adiabatically continuous with product states. iii. Features that are captured in product-state field theories by proliferation of instantons are encoded in perturbative fluctuations about entangled saddles. We develop a general formalism for such path integrals and a couple of simple examples to illustrate their utility. 
\end{abstract}
\maketitle

Feynman's path integral formulation\cite{Feynman:1948qf,item/11282/309393} of quantum mechanics
constructs a description of a system through a weighted sum over classical trajectories --- sequences of classical configurations. Because of this manner of construction, it
 is particularly useful for understanding the emergence of classical behaviour. Extremal classical trajectories satisfying the Euler-Lagrange equations
 dominate the path integral. These trajectories may be understood as proxies for the dynamics of a pure quantum system through its Hilbert space, obtained by continually projecting this dynamics onto a sub-manifold of classical states. Quantum corrections to these classical paths can be found in two ways: by expanding in small fluctuations
 using a Feynman diagrammatic expansion; or by allowing imaginary-time excursions, or instantons, of the dynamics to describe tunnelling processes. Both introduce quantum entanglement into the path integral. 
 
 Entanglement is a defining feature of quantum systems. Focussing upon its structure has led to a new clarity in our understanding of many-body states\cite{Fannes:1992uq,White:1992zl,verstraete2008matrix}. This is perhaps most dramatic in the success of the density matrix renormalisation group\cite{White:1992zl} (DMRG) and its various developments\cite{Schollwock:2005ul}. An important message of DMRG is that the spectrum of entanglement is often the best way to decide which information to retain for an efficient description of a quantum state. Such states may be far from product states and, 
 without this guide, apparently require a large amount of information to describe them. 
 
Quantifying the amount of entanglement is, then, very important. One approach is to use variational states with a bounded degree of entanglement. Matrix product states (MPS) and their higher dimensional analogues do just this\cite{PhysRevLett.98.070201}, with the amount of entanglement being bounded by the rank of the tensors. These states are a very direct realisation of the insights of DMRG\cite{schollwock2011density}. They are restricted superpositions of product ({\it i.e.} classical) states. Although entanglement has been determined from field theories, by evaluating correlation functions in space times with non-trivial geometries\cite{1742-5468-2004-06-P06002}, path integrals are generally not well-suited to this, since they are typically formed from coherent states that are not entangled.

Our aim here is to combine the utility of Feynman path integrals with that of tensor network states.  In order to do this, we construct a path integral as a weighted sum of entangled  trajectories --- trajectories consisting of sequences of weakly entangled, tensor network states. We will refer to these as semi-classical trajectories. The resulting field integral over tensor network states affords several insights.
The extremal trajectories that dominate the path integral correspond to the projection of the Hamiltonian motion through Hilbert space onto the restricted manifold formed by the tensor network states, {\it i.e. } they correspond to the time-dependent variational principle (TDVP) on this manifold\cite{PhysRevLett.107.070601}. These trajectories are semi-classical in two senses.  Firstly, they are described by a number of parameters that scales in a classical, polynomial manner with the size of the system. Secondly, the manifold of tensor network states forms a perfectly good classical phase space and the TDVP can be identified with a classical Hamiltonian dynamics through this phase space\cite{2012arXiv1211.3935H,PhysRevLett.107.070601}. As a result, this field integral can be used to determine when the quantum mechanical degrees of freedom of a complex system conspire to produce emergent semi-classical coordinates. We show how this can be used to provide a complementary picture of deconfined quantum critical points\cite{Senthil:2004qi,Senthil:2004ov}.  
Generally, since the tensor network states are restricted summations of product states, the saddle point of a field integral at a given bond order corresponds to a restricted re-summation of instanton configurations 
at lower bond order.

The entangled path integral should not be confused with continuum tensor networks. Continuum tensor networks\cite{PhysRevLett.104.190405,2012arXiv1211.3935H}  realise an aim of Richard Feynman to construct variational states for quantum fields. They provide a complementary way to combine notions of field theory and tensor networks. The construction is quite different from our entangled path integral, however. The static continuum MPS can be interpreted as a path integral where the direction along the chain is interpreted as a time-like direction --- a holographic dimensional reduction\footnote{As opposed to increase as in MERA or AdS/CFT}. We construct a path integral {\it over} matrix product states\footnote{As opposed to {\it for} the matrix product state in the case of continuum tensor networks.} with a genuine time coordinate and no dimensional reduction. 

The outline of this paper is as follows: In Section~\ref{sec:Construction}, we review the construction of the usual product state path integral and follow with a formal construction of the entangled path integral. Sections~\ref{sec:Geometry} and  \ref{sec:Interpretation} contain formal considerations of the MPS path integral and Section~\ref{sec:ParamandExamples} gives some examples. Depending upon taste, the reader may choose to read Section~\ref{sec:ParamandExamples}  before Sections~\ref{sec:Geometry} and  \ref{sec:Interpretation}. In Section~\ref{sec:Geometry}, we discuss a geometrical interpretation of the dynamical terms in the MPS action. In Section~\ref{sec:Interpretation}, we give physical interpretations of the main features of our results, emphasising the emergent semi-classical coordinates of the saddle points, the connection between the locality of the field theory and efficiency with which the tensor network that describes it can be contracted, and the role of instantons. 
In Section~\ref{sec:ParamandExamples}  we give an explicit parametrisation of the field integral and apply it to two examples --- the one-dimensional $J_1$-$J_2$ model, and a two-dimensional model of a transition from a columnar valence bond solid to a N\'eel state.  Finally, we discuss the broader implications
of our work in Section~\ref{sec.Conclusions}.

\section{Constructing a path integral over entangled states}
\label{sec:Construction}
The conventional Feynman path integral is constructed by following the system through a sequence of product-state field configurations\cite{weinberg1996quantum,altland2010condensed,zee2010quantum}. These configurations are classical in that they harbour no entanglement; they may be, for example, coherent state configurations of the local fields. The sole requirement is that they are dense on Hilbert space, so that any state can be written as a superposition of them and so that a classical limit of smooth trajectories may be taken. 

The field integral --- for example for the partition function --- is constructed by dividing the (imaginary) time evolution operator into many infinitesimal slices and using the over-completeness of the product states to insert resolutions of the identity in the form
${\bm 1} = \int d \psi |\psi \rangle \langle \psi |,$
where $| \psi \rangle $ indicates some product state and $d \psi $ a gauge invariant measure over its parameters. In this prescription, the quantum mechanical structure enters through the overlap of quantum states from one instant to the next\footnote{The states over which the resolution of the identity is constructed must be sufficiently over-complete that the dominant paths can be treated as continuous in time. This is not the case for example if coherent states are restricted to a great circle of the Bloch sphere. The states that we use satisfy this requirement.}. The partition function is given  by\cite{feynman1998statistical,nagaosa2013quantum}
\begin{eqnarray}
{\cal Z} 
= \mbox{Tr}\, e^{- \beta {\cal H}}
= \int D\psi(\tau) e^{-{\cal S}[\psi]}.
\label{ProductPartition}
\end{eqnarray}
The trajectories are weighted with an action composed of a dynamical (or Berry phase\cite{berry1985classical}) term describing this overlap and the expectation of the Hamiltonian;
$
{\cal S}[\psi]=\int_0^\beta d\tau \left[ \langle \psi | \dot \psi \rangle -\langle \psi| \hat {\cal H} | \psi \rangle \right].
$
The crucial point in this construction is that the states over which we resolve the identity should be over-complete, so that an unbiased measure may be written over them and a classical limit of smooth trajectories taken. 

We extend the Feynman path integral construction in one dimension by following the system through a sequence of entangled states. For these, we choose MPS states. These are an extension of product states where the local coefficients are endowed with auxiliary indices that are contracted on the spatial network --- a line in the case of MPS states:
\begin{eqnarray}
 | \psi_A \rangle 
&= &
\sum_{\{ \sigma_i \},\{a_i\},...} V^L_{a_0}
... A_{a_{i-2}a_{i-1}}^{\sigma_{i-1}}A_{a_{i-1} a_{i}}^{\sigma_{i}}A_{a_{i} a_{i+1}}^{\sigma_{i+1}} 
... V^R_{a_N}  
\nonumber\\
& & \;\;\;\;\;\;\;\;\;\;\;\;\;\;\;\;\;\;\;\;\;\;\;\;
\times
|\sigma_{i-1} \rangle \otimes |\sigma_{i} \rangle  \otimes |\sigma_{i+1} \rangle...  
\nonumber\\
&=&
\raisebox{-0.25in}{\includegraphics[width=2.5in]{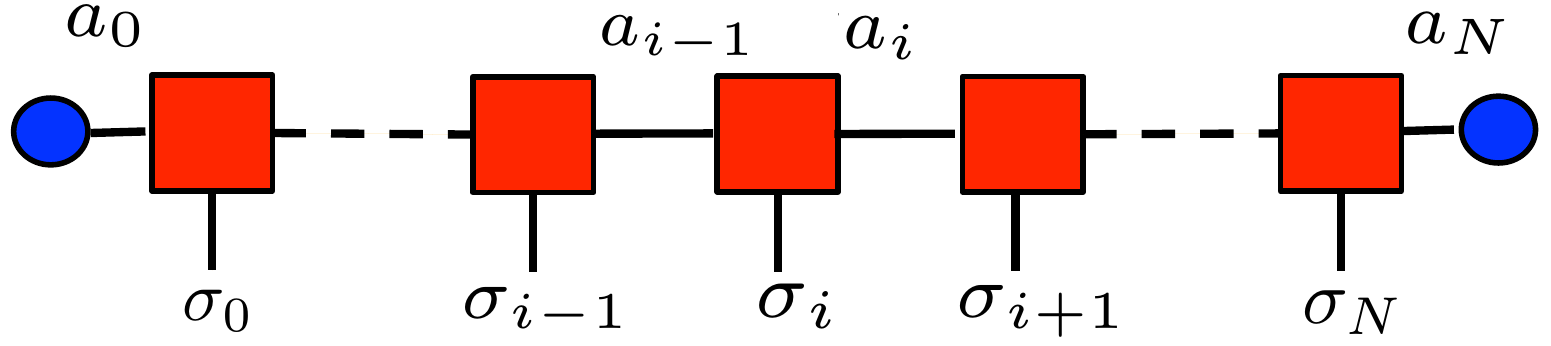}}
\label{MPSDefn}
\end{eqnarray}
Here, $\{|\sigma_i \rangle  \}$  form a basis of spin states 
(we consider only the spin half case) on the site $i$. The tensors $A^\sigma_{ab}$ have dimension $d \times D\times D$, where $d$ is the dimension of the local Hilbert space ($2$ in the case of spin half) and $D$ is the dimension of the auxiliary indices --- often called the bond dimension. The auxiliary vectors $V^L_a$ and $V^R_a$ terminate the chain. Often the state is independent of them in the thermodynamic limit, although when translational invariance is broken ({\it e.g.} by different dimer coverings) there may be residual dependence. Eq.(\ref{MPSDefn}) also shows a common graphical representation of this state. 

The entangled field integral can be constructed by expanding the time evolution operator with the insertion of resolutions of the identity over MPS states at successive times:
${\bm 1} = \int D A |\psi_A \rangle \langle \psi_A |.$
That such a resolution of the identity can be made is guaranteed by the over-completeness of the MPS states\footnote{This over-completeness itself follows from the fact that the MPS states determined by Eq.(\ref{MPSDefn}) are restricted summations over product state, which are themselves over-complete}. We shall discuss an appropriate choice of the measure, indicated symbolically by $DA$, presently. 
Armed with this resolution of the identity over weakly-entangled states, we can construct a field integral for the partition function precisely as before, by inserting resolutions of the identity between infinitesimal time evolutions. The resulting expression for the partition function is identical to Eq.(\ref{ProductPartition})
with $|\psi \rangle  \rightarrow | \psi_A \rangle$, and the measure represented by $DA$.

A gauge invariant measure  for the MPS field integral
can be constructed in the following manner: First we write the MPS as a matrix product operator (MPO) acting on some reference state --- a product state over spin-up states say. 
To achieve this we write $A^\sigma_{ab}= {\cal A}^{\sigma \delta}_{ab} v^\delta$, where $v^\delta$ are the spinor components of the reference state: $| {\bf v} \rangle = v^\delta |\delta \rangle$. Next, we pair the indices of each  $D \times d \times D \times d$-dimensional tensor of this MPO according to ${\cal A}^{(a,\sigma)}_{(b,\delta)}$. Written in this way, the tensor can be interpreted as a quantum circuit that, moving from right to left along the chain, sequentially makes unitary maps from the combined reference spin-space and auxiliary space to the physical spin-space and auxiliary space at the next site. ${\cal A}^{(a,\sigma)}_{(b,\delta)}$ is therefore an element of $SU(dD)$ for which we can construct a Haar measure\footnote{Pairing the indices in this way, and interpreting the MPO as a circuit that contracts from right to left, is a gauge choice. We could have made the opposite choice $A^{(b,\sigma)}_{(a,\delta)}$ which would have led to an interpretation as a circuit acting from left to right along the chain. The gauge choice that does not change the physical state.}.  Without loss of generality, we may choose ${\bf v}$ so that only its first component is non-zero and we shall do this from now on.

\begin{figure}[t!]
 a)
\raisebox{-1in}{\includegraphics[width=2.8in]{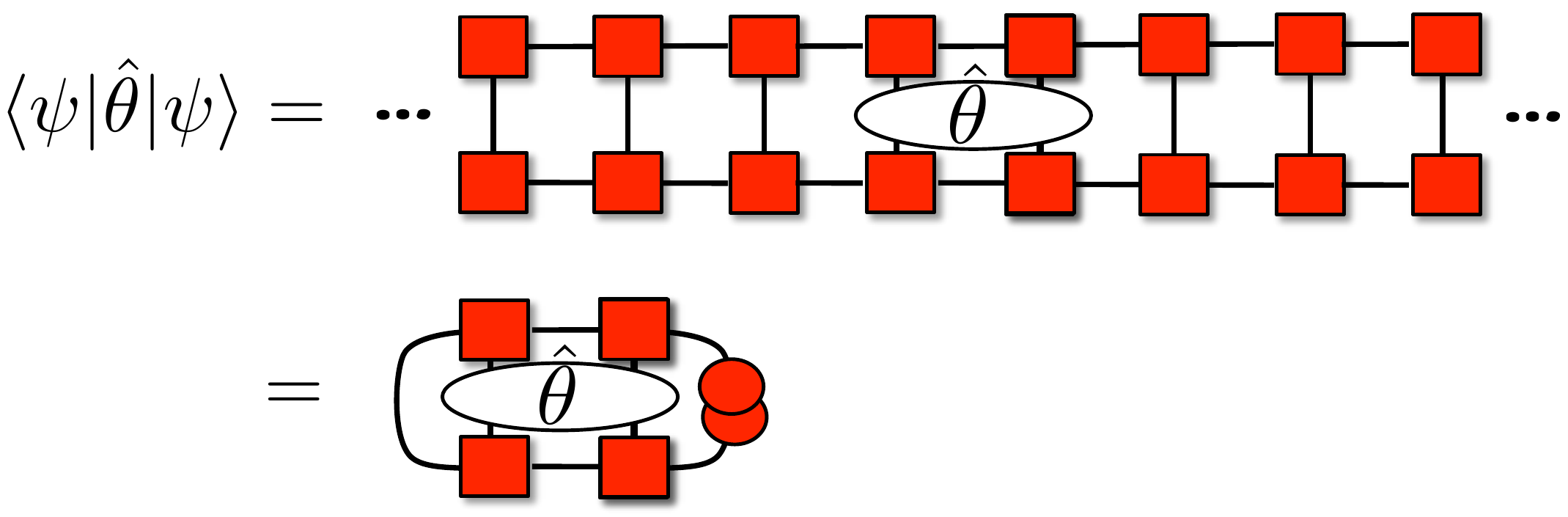}}\\
\hspace{-0.8in} 
b)
\raisebox{-0.8in}{\includegraphics[width=2in]{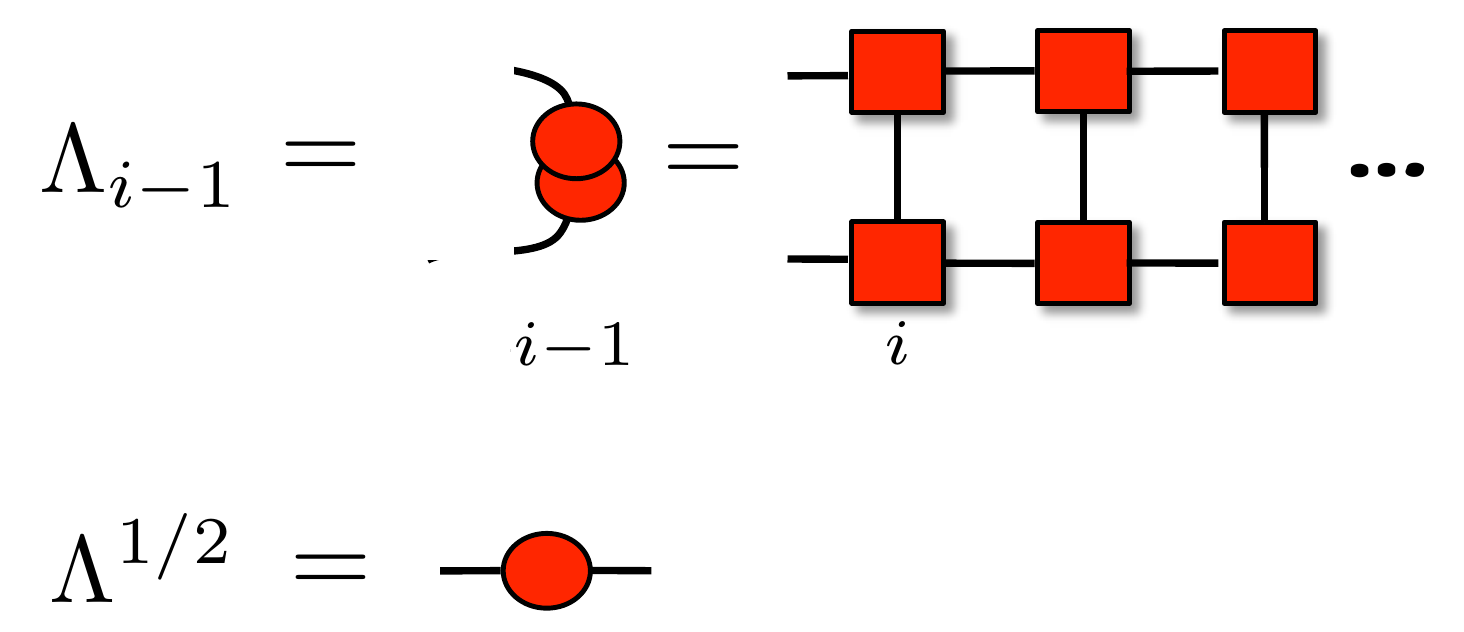}}
\begin{center}
\caption{{\it Graphical representation of the non-locality of MPS expectations and definition of the environment tensor.} a) The expectation of a local operator is generally non-local for an MPS because of entanglement. It is local when expressed in terms of both the tensor $A^\sigma_{ab}$ and the environment $\Lambda_{ab}$. b) The environment at the site $i$ is given by the contraction of the tensor network to the right of that site. It is also useful to define the square root of this. We choose a gauge in which the contraction to the left gives the identity. }
\label{fig:RightEnvironment}
\end{center}
\end{figure}

The locality of the field integral is an important requirement if it is to be useful. This is guaranteed for product states, for which expectations of local operators are local. For MPSs, the expectation of a local operator requires a contraction along the whole chain. {\it A priori}, then, the MPS action 
is not local. This can be readily circumvented by introducing additional local degrees of freedom into the field integral that describe the contraction of the MPS to the left and right of any local operator [see Fig. \ref{fig:RightEnvironment}]. In fact, we have already made an implicit choice of gauge for our MPO tensors so all we have to keep track of is the contraction (or environment) to the right of a given point in the chain. This is a $D \times D$-dimensional tensor, $\Lambda_i$, given by the equation
\begin{eqnarray}
\Lambda_{i-1} 
= \sum_\sigma {{\cal A}_{i}^{\sigma,1}} \Lambda_{i} {{\cal A}_{i}^{1, \sigma}}^\dagger
\;\;\;\; \hbox{or} \;\;\;\;
\raisebox{-0.24in}{\includegraphics[width=0.8in]{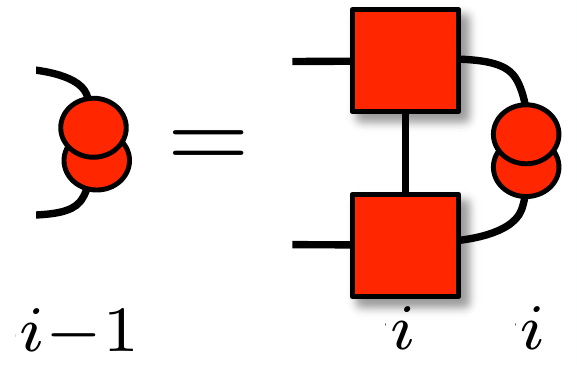}}
\label{Environment}
\end{eqnarray}
where we have suppressed auxiliary indices for clarity; each term is a matrix in auxiliary indices and the contraction is simply that of usual matrix multiplication. With this definition of the environment, the field integral may be written as 
\begin{eqnarray}
{\cal Z} 
&=&
 \int D{\cal A} D\Lambda 
 \,
\delta \left[  \Lambda_{i-1}
- \sum_\sigma {{\cal A}_i^{\sigma,1}} \Lambda_{i} {{\cal A}_i^{1,\sigma}}^\dagger
  \right] e^{-{\cal S}[{\cal A}]}.
  \label{MPSPartition}
\end{eqnarray}
The integral over the environment, $\Lambda$, may be constructed freely over its components, imposing the additional normalisation constraint $\mbox{Tr}\, \Lambda=1$ on the first link of the chain. Eq.(\ref{Environment}) guarantees that this constraint is satisfied on all other sites due to the unitarity of ${\cal A}$. This is now a conventional --- though complicated --- path integral that may be manipulated by standard means.
%

\section{The Geometry of the Berry Phase}
\label{sec:Geometry}
The dynamical Berry phase\cite{berry1985classical}  encodes the role of quantum mechanics in the Feynman path integral. It is determined by the overlap of infinitesimally separated points on the variational manifold. In the case of a single spin half, or a product state of spins-half coherent states, it has an appealing geometrical interpretation. One familiar form is\cite{auerbach2012interacting,tsvelik2007quantum}
\begin{equation}
{\cal S}_B
= \frac{1}{2} \int dt {\bf A} [{\bf n} ] . \dot {\bf n},
\label{WessZumino}
\end{equation}
where ${\bf n}$ is an $O(3)$ vector parametrising the spin coherent state, and ${\bf A} [{\bf n} ]$ is the vector potential for a monopole placed at the centre of the $O(3)$ sphere. For periodic paths in real time it is proportional to the solid angle swept out by the spin on the Bloch sphere. This geometrical interpretation is very powerful, leading ultimately to a geometrical picture of the quantisation of spin.  The Berry phase for MPS has a similar geometrical interpretation. In order to reveal this, we build up gradually to a general expression. 

\vspace{0.02in}
{\it Berry Phase for Small Paths:} The first case is that of trajectories that deviate only slightly from some reference point on the variational manifold. As a warm up to this, consider the case of a product state over spin coherent states, $| {\bf N} \rangle= | {\bf n}_1, {\bf n}_2, ... \rangle$. A trajectory that deviates only slightly from some particular product state can be followed by expanding on each site according to 
${\bf n}= {\bf n}_0 (1-{\bf l}^2 )^{1/2} + {\bf l}$, with $ {\bf n}_0.{\bf l}=0$. 
The rotation from $|{\bf n}_0 \rangle$ to $|{\bf n} \rangle$ can be achieved using the operator 
$\exp [ \frac{i}{2}   {\bf n}_0 \times {\bf l} \cdot {\bm \sigma}]$, which rotates by an angle $|{\bf l}|$ about the axis ${\bf n}_0 \times {\bf l}$. To leading order in ${\bf l}$, we may write 
$|{\bf n} \rangle = |{\bf n}_0 \rangle+ |d{\bf n}_0 \rangle$, 
 with $|d{\bf n}_0 \rangle= \frac{i}{2} ( {\bf n}_0 \times {\bf l} \cdot {\bm \sigma} ) | {\bf n}_0 \rangle $
 so that 
 $\langle {\bf n}_0 | d{\bf n}_0 \rangle =0$. 
 The small change $| d{\bf n}_0 \rangle$ in the state $| {\bf n}_0 \rangle$ is a tangent to the manifold of coherent states. 
 The Berry phase in this case takes the rather simple form 
 \begin{equation}
 {\cal S}_B= \int dt \langle {\bf N} | \dot {\bf N} \rangle = \sum_i \int dt  \; l_{(i)}^* \dot l_{(i)} ,
 \end{equation}
where $l=l_1+il_2$, and $l_1$ and $l_2$ are two components of ${\bf l}$ for each site.

The analogous case for MPS states takes a similar form. We consider small paths about some reference state, the tensor on the site $i$ being given by $A^\sigma_{ab(i)}(x)=A^\sigma_{ab(i)}+dA^\sigma_{ab(i)}(x)$. A tangent vector is obtained by replacing $A^\sigma_{ab}$ at the site $i$ in Eq.(\ref{MPSDefn}) by $dA^\sigma_{ab(i)}$, where  $dA^\sigma_{ab(i)}$ is chosen so that the resulting state is orthogonal to the original MPS. The $dA^\sigma_{ab(i)}$ satisfying this constraint can be parametrised by  $D \times (d-1)\times D$ tensors $x^{\sigma\ne1}_{ab(i)}$at each site\cite{PhysRevLett.107.070601}  --- these play an analogous role to ${\bf l}_{(i)}$. The variations of the tensor at a particular site (we suppress the site index for clarity) are given explicitly in terms of the MPO by
\begin{eqnarray}
dA^\sigma_{ad}(x) 
&=&
{\cal A}^{\sigma,\delta \ne 1}_{ab} x^{\delta \ne 1}_{bc} {\Lambda}^{-1/2}_{cd},
\nonumber\\
&=&
\raisebox{-0.3in}{\includegraphics[width=0.9in]{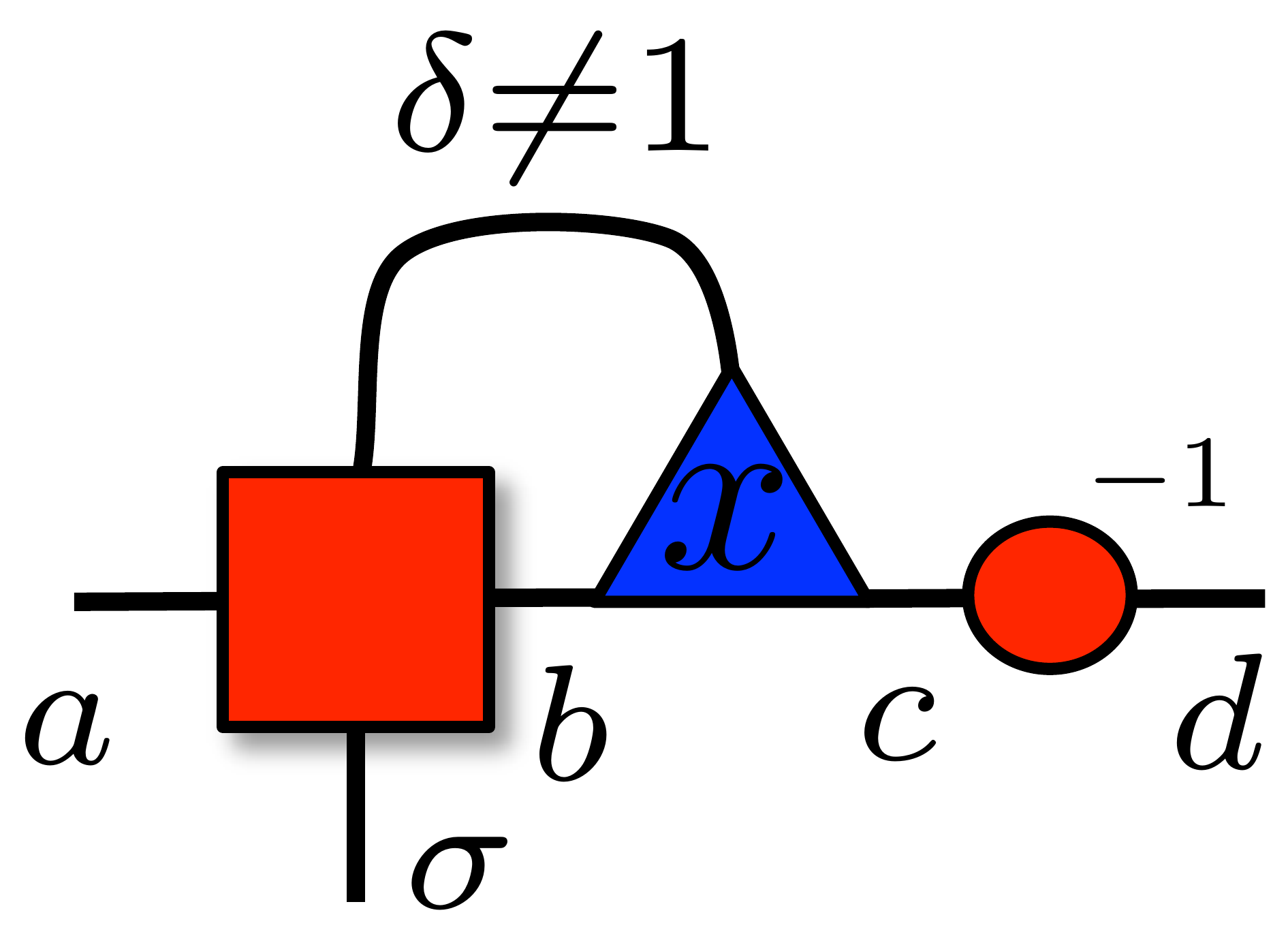}}
\label{MPStangentvec}
\end{eqnarray}
where the graphical representation has the interpretation laid out in Eq.(\ref{MPSDefn}) and Fig.\ref{fig:RightEnvironment}. Notice that the MPO contains both the reference MPS and the structure of the tangent space. 
The Berry phase  for such a trajectory can be written as
\begin{equation}
{\cal S}_B= \sum_i \int dt \; \sum_{\sigma \ne 1} \mbox{Tr}\, [{x^{\sigma}_{(i)}}^\dagger \dot x^{\sigma }_{(i)}].
\label{MPSBerryIII}
\end{equation}
Auxiliary indices of $x^\sigma_{ab}$ have been suppressed in favour of a matrix notation. 
This construction builds upon the time-dependent variation principle, introduced in the context of MPS states by Verstraete {\it et al.}\cite{PhysRevLett.107.070601}. Amongst the insights of 
Ref.~\onlinecite{PhysRevLett.107.070601} was a choice of gauge for the tangent vectors such that their overlaps are zero for variations on spatially separated sites and, moreover, such that the tangent vectors form an orthonormal set for variations on the same site\footnote{
The details of this are best found in Ref.\cite{PhysRevLett.107.070601}. In the gauge chosen in Ref.\onlinecite{PhysRevLett.107.070601} and the notation that we have adopted here, the variation of the MPS tensor on a given site is Eq.(\ref{MPStangentvec}).
In the notation of Haegemann {\it et al.}, 
$V^{(\sigma a)}_{\beta\equiv(\delta, c)}={\Lambda_L^{1/2}}_{ab}   A^{\sigma,\delta \ne 1}_{bc}$
 is a matrix constructed from the $D(d-1)$, $dD$-dimensional null-vectors of the matrix $({A^\sigma}^\dagger \Lambda^{1/2}_L)_{ab}$ [after reshaping the latter to form a $D \times dD$-dimensional matrix by regrouping the indicies as  $({A}^\dagger \Lambda^{1/2}_L)_{a, (\sigma,b)}$]. 
\\
Although apparently complicated, this prescription for constructing the tangent vectors leads to a dramatic simplification. The tangent vectors have zero overlap if they correspond to variations on different sites, and form an orthonormal set on site. The overlap or Gramm matrix of two tangent vectors parametrised by variations, $ dA^\sigma_{ab}(x)$ and $dA^\sigma_{ab}(y)$, of MPS the tensors on the same site is given by  
$ \sum_{\sigma \ne 1}Tr [{x^{\sigma}}^\dagger y^{\sigma }]$.
}. The essence of this is similar to the spin-wave expansion where the Bloch sphere is treated as locally flat. 

\vspace{0.02in}
{\it A general expression for the MPS Berry phase } can be constructed using an explicit parametrization of the full MPS manifold.  Wouters {\it et al.}\cite{wouters2013thouless} 
obtained a compact parametrization by using the Thouless theorem to effectively exponentiate Eq.(\ref{MPStangentvec}). We use an equivalent parametrization  that emphasises a connection with the spinor representation of spin coherent states [see Appendix \ref{App: Parametrising MPS states}]:
\begin{eqnarray}
A^\sigma_{ad(i)}(z)
&=&
{\cal A}^{\sigma \delta}_{ab(i)} z^\delta_{bc(i)} \Lambda^{-1/2}_{cd(i)} 
=
\raisebox{-0.25in}{\includegraphics[width=1in]{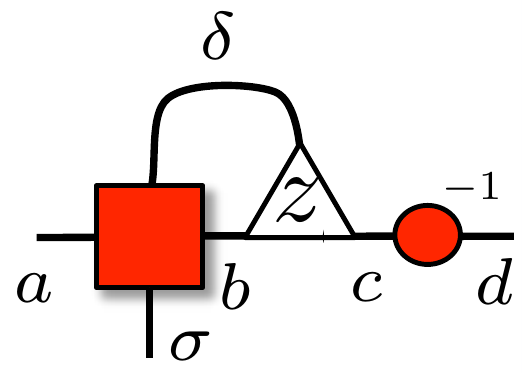}}
,
 \label{SpinorMPS}
\end{eqnarray}
with consistency
\begin{eqnarray}
& & 
 \sum_{\sigma=1}^d \mbox{Tr}\, [ {z^\sigma}^\dagger z^\sigma ]  
 =\raisebox{-0.27in}{\includegraphics[width=0.45in]{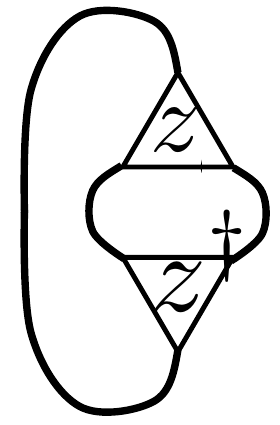}}
 =1,
 \nonumber  \\
 & &
 \Lambda_{(i)} 
=
  {z^\dagger}^\sigma_{(i)} z^\sigma_{(i)} = {\cal A}^{\sigma \delta}_{(i+1)} z^\delta_{(i+1)} {z^\dagger}^\gamma_{(i+1)} {{\cal A}^\dagger}^{\gamma \sigma}_{(i+1)}
  \nonumber\\
  & &
  \raisebox{-0.27in}{\includegraphics[width=2.5in]{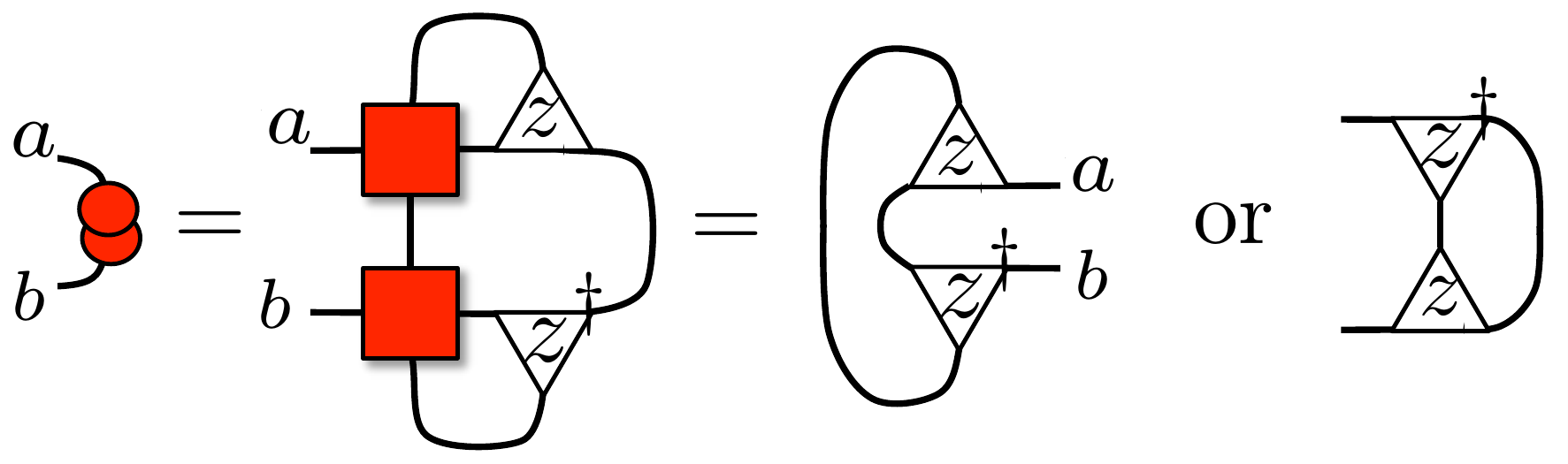}}.
\label{SpinorMPSconsistency}
\end{eqnarray}
The first of these imposes normalisation of the MPS state and the second encodes consistency between contracting the transfer matrix to the left or right. We have suppressed auxiliary indices for clarity, and the subscripts indicate the lattice site.

The form of Eq.(\ref{SpinorMPS}) suggests an intriguing interpretation. We have already commented that the MPO contains both the MPS in its components ${\cal A}^{\sigma,1}_{ab}$ and the tangent vectors at this point in its other components. In fact, the MPO encodes $d$ orthonormal MPSs in the set of tensors 
$\{ {\cal A}^{\sigma,1}_{ab}, \; {\cal A}^{\sigma,2}_{ab}\; ... \; {\cal A}^{\sigma,d}_{ab} \}$.
Eq.(\ref{SpinorMPS}) is a generalisation of the spinor representation of a spin coherent state, with 
the $d\times D \times D$ tensors $z^\sigma_{ab}$ the generalised spinor components weighting these orthogonal states. In the case of $D=1$, Eq.(\ref{SpinorMPS}) reduces to the ${\cal C} P_1$ representation of the spin coherent state on the site $i$. 

Using Eq.(\ref{SpinorMPS}), a trajectory on the MPS manifold may be followed by a time-dependent set of parameters, $z^\sigma_{ab} (t)$, together with a static reference MPO ${\cal A}^{\sigma,\delta}_{ab}$. The Berry phase of a trajectory described in this way is given by
\begin{equation}
{\cal S}_B= \sum_i \sum_\sigma  \int dt \;
\mbox{Tr}\,
[{z^\sigma_{(i)}}^\dagger \dot z^\sigma _{(i)}],
\label{MPSBerryII}
\end{equation}
which for bond order, $D=1$, reduces to the spinor representation of the coherent state Berry phase Eq.(\ref{WessZumino}),
with $z$ is related to ${\bf n}$ by a Hopf map; ${\bf n}= z^\dagger {\bm \sigma} z$. Eq.(\ref{MPSBerryII}) gives an appealing local construction of the Berry phase.

\section{Interpreting the Entangled Path Integral}
\label{sec:Interpretation}

{\it Restricted paths are classical:} A finite bond-order MPS forms a classical representation of a quantum state, in the sense that it requires an amount of information that scales linearly with the length of the system rather than exponentially. This interpretation of the MPS as a semi-classical state can be taken further by noting that the dynamics projected onto the manifold of such states can be associated with a classical Hamiltonian dynamics\cite{PhysRevLett.107.070601,2012arXiv1210.7710H}. Introducing the entanglement variables of the MPS into a field integral can be interpreted as the emergence of a new set of semi-classical, collective coordinates.  We use the term semi-classical to distinguish this from products of coherent states, which have no entanglement and are more conventionally referred to as classical. 

{\it Adiabatic Continuity:} In what circumstances is it necessary to use a higher bond-order field theory to describe a system? If our analysis is to be perturbative, the saddle point of the low bond order field theory should be adiabatically continuous with the actual saddle-point configuration. This concept is familiar in Fermi liquid theory where it refers to the preservation of the quantum numbers of a state --- the number of quasi-particles {\it etc.} --- while its qualitative features may be changed markedly. This may be translated to the language of variational manifolds as follows: an arbitrary state in the Hilbert space can be represented as a sum of states, or wavepacket, on the variational manifold (the variational manifold is given by the states of the Fermi gas in the case of Fermi liquid theory). If this wavepacket forms a connected region then it may be constructed by perturbative corrections from a representative point in the region --- albeit possibly requiring a very high order. If the wavepacket is disconnected, however, it cannot be constructed perturbatively and requires tunnelling between the patches to capture its physical effects. The simplest example of this is the triplet state formed by anti-symmetric superposition of states $|{\bf l}, -{\bf l} \rangle $ and $|-{\bf l}, {\bf l} \rangle $ --- states that are diametrically opposite on the variational manifold of product states. In extreme cases, the different patches that comprise the wavepacket may harbour topological differences in the states that they describe. 

Remarkably, even in situations where entanglement diverges --- such as at a quantum critical point\cite{sachdev2011quantum} --- a product state saddle point provides an adequate starting point for a field theoretical analysis. When this is not the case, such as at deconfined critical points and cases where the physics is driven by non-perturbative effects, a higher bond-order field theory may be required as a good starting point. The spirit here is to find the lowest bond order that adiabatically connects the saddle to the actual groundstate. This is slightly different from the usual spirit of MPS, where bond order is increased until a required degree of numerical convergence is achieved. In many cases, the required bond order might turn out to be rather low\cite{crowley2014quantum}.

\vspace{0.1in}
{\it Instantons and entangled paths:}
The idea of adiabatic continuity  is related to potentially one of the most useful features of the entangled path integral. Tunnelling involves a system passing through an intermediate superposition of two or more classical states. Such configurations are not represented by real-time saddle point configurations of a product state field theory. Instead, imaginary time excursions or instantons are required. However, since MPSs are constructed explicitly as a (restricted) sum of product states, real time trajectories that transfer weight from one classical trajectory to another are possible. In this way, a field integral constructed at a given bond order resums instantons of the lower bond order or product state theory. We give an explicit example below where key physics is driven by the proliferation of instantons in the product state theory --- the same physics is captured in the bare saddle point of the bond order three theory. It is intriguing to speculate that this may be related to the observation of resurgence in field theory\cite{dunne2012resurgence,dunne2013continuity,dunne2014generating}  --- a connection between perturbative and instanton contributions to certain field integrals. 

\vspace{0.1in}
{\it Efficient contractability vs locality of field integral:} In Sec.~\ref{sec:Construction}, we showed that the construction of a local path integral with a finite number of fields at each point places important constraints upon the class of entangled states over which the field integral can be constructed. In particular, we require that the result of contracting the tensor network in the region surrounding the support of some local operator  --- the environment of this region --- can be summarised by a finite amount of data. This is always the case for MPS states. Extensions to higher dimensions are trickier\cite{verstraete2008matrix} --- though possible.

\section{An Explicit Parametrization and Some Simple examples}
\label{sec:ParamandExamples}
After the formal construction of the entangled path integral in Sec.~\ref{sec:Construction}, ~\ref{sec:Geometry} and ~\ref{sec:Interpretation}, we now apply it to two simple examples chosen to illustrate the new features highlighted in Sec. \ref{sec:Interpretation}. We show how fluctuations about a low bond order saddle point generate physics that cannot be captured in either a product state field theory 
or in the low bond order saddle point equations. Indeed, any entanglement in the path integral enables new physics to be captured beyond that of the conventional path integral. Therefore, in order to keep the analytical complexities to a minimum whilst retaining key physics, we consider the  following restricted bond-order three MPS state:
\begin{eqnarray}
|A_i\rangle
&=&
\sum_\sigma A_i^\sigma |\sigma \rangle
=
\left(
\begin{array}{ccc}
0 & | {\bf l}_i \rangle & | - {\bf l}_i \rangle \\
n_{1,i-1}| {\bf l}_i \rangle & 0 & 0 \\
n_{2,i-1}| - {\bf l}_i \rangle & 0 & 0 
\end{array}
\right)
\label{SimpleMPS}
\end{eqnarray}
%
%
which parametrizes magnetic and local singlet order in a 
similar manner to the bond operators of Ref.\onlinecite{sachdev1990bond}, but without breaking translational symmetry. In Eq.(\ref{SimpleMPS}), $|{\bf l}_i\rangle $ is a spin coherent state on site $i$ parametrised by the $O(3)$ vector ${\bf l}_i$, and $|-{\bf l}_i \rangle$ is orthogonal to it. $n_i=(n_{1,i},n_{2,i})$ is an $SU(2)$ spinor satisfying the relation $n_i^\dagger n _i= |n_{1,i}|^2+ |n_{2,i}|^2=1$, which 
characterises 
entanglement across the bond between site $i$ and site $i-1$. This is revealed by the environment tensor - the result of contracting the MPS to the right\footnote{The contraction to the left gives the identity.} of a site $i$; 
\begin{eqnarray}
\Lambda_i = \hbox{diag} \left[ \lambda_{i+1},\lambda_{i} |n_{1,i}|^2 , \lambda_{i} |n_{2,i}|^2 \right],
\label{environment}
\end{eqnarray}
where  $\lambda_i= \lambda $ for $i$ even and $ \lambda_i= 1-\lambda $ for $i$ odd. The additional global parameter $\lambda$ gives the relative weight of the two different singlet coverings of the line\footnote{and summarises a residual dependence upon the terminating auxiliary vectors $V_L$ and $V_R$ mentioned after Eq.(\ref{MPSDefn})}.

The Berry phase for such states is given by
\begin{eqnarray}
{\cal S}_B &=&
\sum_i
\int dt 
\left[
\lambda_i \langle {\bf n}_{i} | \dot {\bf n}_{i} \rangle
+( \lambda_{i-1}n_{i-1}^z  +\lambda_i n_{i}^z ) \langle {\bf l}_{i} | \dot {\bf l}_{i} \rangle  \right],
\nonumber\\
\label{SimpleMPSBerry}
\end{eqnarray}
where $\langle {\bf l}_{i} | \dot {\bf l}_{i} \rangle$ indicates the usual Berry phase for a spin coherent state. $\langle {\bf n}_{i} | \dot {\bf n}_{i} \rangle$ indicates the same for the spinor $n$, expressed in terms of the $O(3)$ vector ${\bf n}$ obtained from $n$ by a Hopf map [we use the same symbol for the spinor and $O(3)$ vector - it being clear from context which is intended; {\it e.g.} $n_x$ and $n_z$ indicate the $x$- and $z$-components of the $O(3)$ field ${\bf n}$]. 

This form of the Berry phase can be deduced by first constructing the $z$-tensors from Eq.(\ref{SimpleMPS}).  Using Eq.(\ref{SpinorMPS}) with the trivial choice of reference MPO
${\cal A}^{\sigma \delta}_{ab} = \delta_{\sigma \delta} \delta_{ab}$, we write 
$z^\sigma_{ac}= A^\sigma_{ab} \Lambda^{1/2}_{bc}$.
%
This satisfies the consistency equations (\ref{SpinorMPSconsistency}) with the environment given by Eq.(\ref{environment}). The Berry phase is then obtained using Eq.(\ref{MPSBerryII}) [see Appendix \ref{App: Parametrising MPS states}].
Finally, the path integral Eq.(\ref{MPSPartition}) may be constructed with the usual Haar measures for ${\bf n}$ and ${\bf l}$, and an integration over the global parameter $\lambda$. Note that no integration over the environment tensor nor functional delta function is required, since Eq.(\ref{SimpleMPS}) allows us to construct the environment tensor, Eq.(\ref{environment}),  without further restriction. 

\subsection{Columnar VBS to Neel transition}
\begin{figure}[t!]
\begin{center}
\includegraphics[width=\linewidth]{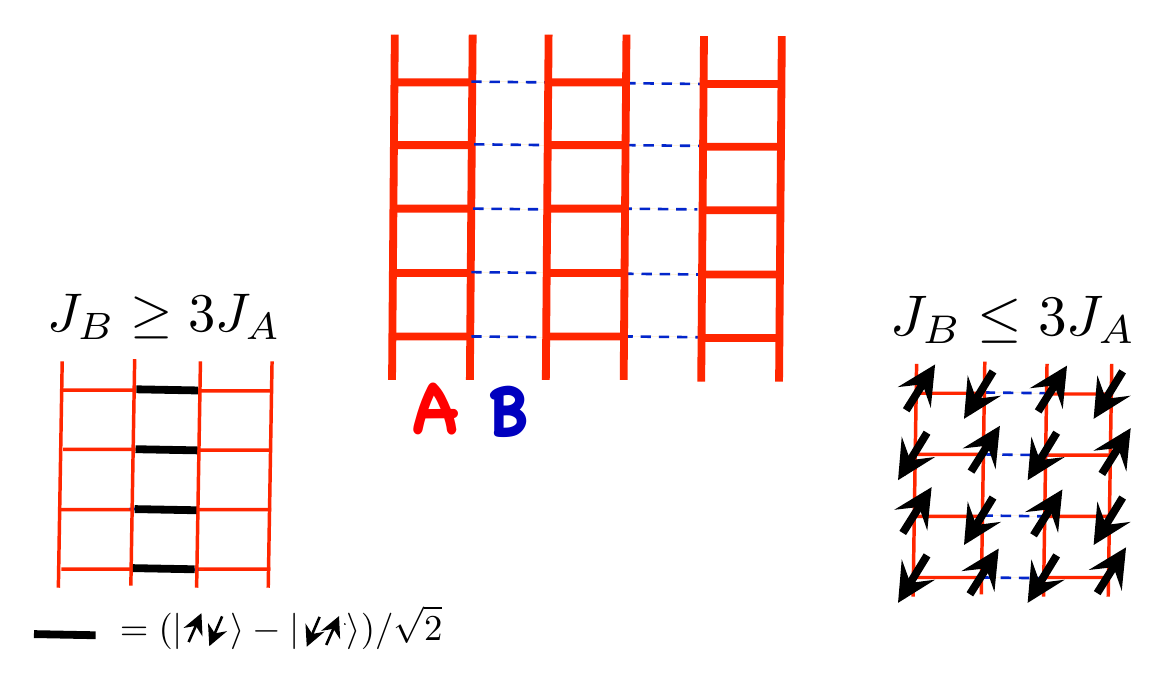}
\caption{{\it Anisotropic antiferromagnetic couplings on a square lattice:} The couplings on the sub-lattices A (red) and B (blue) are different. The system displays a phase transition from a columnar lattice of singlets on the B lattice when $ J_B \gtrsim 3J_A$ to a N\'eel state when $ J_B \lesssim 3 J_A$. }
\label{fig.VBStoNeelLattice}
\end{center}
\end{figure}

The first application that we consider concerns a transition between a columnar valence bond solid and N\'eel order. 
This transition has no conventional Ginzburg-Landau description and displays some of the key features of a deconfined quantum critical point\cite {Senthil:2004qi}, although the model that we consider explicitly breaks translational symmetry.
It is a revealing, simple example of the insights that may be gained from the MPS path integral. 
The Hamiltonian is simply an antiferromagnetic Heisenberg model,
${\cal H} = \sum_{\langle ij \rangle} J_{ij } {\bm \sigma}_i. {\bm \sigma }_j $,
with  different coupling strengths on two sub-lattices. As shown in Fig.~\ref{fig.VBStoNeelLattice}, these sub-lattices correspond to the legs and rungs of a series of parallel ladders --- sub-lattice A --- and additional rungs that connect them --- sub-lattice B. 
The system undergoes a quantum phase transition as $J_A/J_B$ is varied  between a 
N\'eel state, when $3 J_A \gtrsim  J_B$, and a valence bond solid of singlets on the B lattice, when $3 J_A \lesssim  J_B$. 

Since this is a two-dimensional lattice, we must use a two dimensional tensor network to describe it. However, as the singlet correlations develop preferentially in the x-direction, we may take a two-dimensional PEPS with bond order 1 in the vertical direction as a leading approximation; {\it i.e. } a state consisting of MPS in the x-direction and no entanglement in the vertical direction. 

The simple bond-order $3$ parametrisation of Eq.(\ref{SimpleMPS}) captures the key physics of the two phases: the Ne\'el state  is obtained with $\lambda=0$ or $1$, $n^z=1$ and ${\bf l}_i=-{\bf l}_{i+1}$; the singlet phase by $\lambda =1$, $n_1=n_2=1/\sqrt{2}$ and ${\bf l}_i=-{\bf l}_{i+1}$.
In general, we expect a mixture of singlet and triplet configurations, as well as singlets on the A-legs that are not described by the states Eq.(\ref{SimpleMPS}).
Notice that the spin coherent state parts of the tensors for the two phases are the same. Evidently, the critical point between the two is a transition in entanglement, hence the usual description in terms of an emergent gauge field. Here we construct a field theory for the critical region explicitly in terms of an entanglement field $n$. 

The saddle point of our MPS path integral is found by optimising the expectation of the Hamiltonian with the state given by Eq.(\ref{SimpleMPS}) over its various parameters. This optimisation fixes $\lambda=1$ 
and ${\bf l}_{i,j}=-{\bf l}_{i+1,j}=-{\bf l}_{i,j+1}$ ($i$ labelling the direction along the MPS and $j$ perpendicular to it) $\forall J_A$  [see Appendix \ref{App: Columnar VBS to Neel Transition}]. The residual dependence upon $n$  is given by 
$$
{\cal H}/N = -3 J_A ({n^z})^2 - 2 J_B {n^x} -J_B ,
$$
which has minima at 
\begin{eqnarray*}
n^z&=&0 \;\;\;\;\;\;\;\;\;\;\;\;\;\;\;\;\;\;\;\;\;\;\;\;\;\;  \hbox{     for } J_B >3 J_A \\
n^z&=&\sqrt{1-J_B^2/(9 J_A^2)}    \;\;\;   \hbox{  for } J_B < 3 J_A.
\end{eqnarray*}
These results suggest a continuous transition in the entanglement structure at $J_B=3 J_A$. 

Critical fluctuations involve going beyond the saddle point MPS to determine the effective theory for fluctuations about it. Since the transition is a transition in the entanglement structure, the important fluctuations are entanglement fluctuations. These can be captured from the effective theory of the field ${\bf n}$. To leading approximation this is given by the  two-dimensional transverse-field Ising model with easy axis anisotropy:
\begin{eqnarray*}
{\cal H} &=&
-J_A \sum_{i',j} \left[ n^z_{i',j} n^z_{i'+2,j}+n^z_{i',j} n^z_{i',j+1}+({n^z_{i',j}})^2 \right]
\\
& & 
\;\;\;\;\;\;\;\;\;\;\;
+2 J_B \sum_{i'j} n^x_{i'j},
\end{eqnarray*}
where $i'=2i$ labels even sites only - the spinor $n$ on odd sites of the lattice drops out of the Hamiltonian for $\lambda=1$. This Hamiltonian will be substantially renormalised by fluctuations of the Ne\'el order, which diverge as $n^z \rightarrow 0$ at the critical point [see Appendix \ref{App: Columnar VBS to Neel Transition}] and ultimately may drive the transition first order\cite{kuklov2008deconfined,lou2009antiferromagnetic,sandvik2010continuous}.

The lesson of this simple model is that an MPS field can provide an order parameter for unconventional phase transitions. The transition is one of entanglement structure, whose usual description is of a non-Ginzburg-Landau type in terms of an emergent gauge field. The dual description provided here is of Ginzburg-Landau type in terms of the MPS field and critical fluctuations are explicitly fluctuations of entanglement structure. 

The possibilities are wider.  In more conventional situations, fluctuations may drive transitions to new phases that would not be found in a mean field analysis. Next, we will study a situation in which fluctuations drive a change in entanglement structure by a mechanism akin to Villain's order-by-disorder;
an effect that would require a condensation of instantons in the conventional theory.

\subsection{$J_1$-$J_2$ Model}
\label{Sec:J1J2Model}
As argued above, an MPS field theory can capture processes in its saddle point configurations that require a proliferation of instantons in a conventional field theory. Our next example concerns a model in which a qualitative change in behaviour is driven by precisely such processes.
In the MPS field theory, the same effect is driven by a Villain order-by-disorder mechanism where quantum fluctuations 
 favour a change in entanglement structure. Consider the $J_1$-$J_2$-model\cite{nersesyan1998incommensurate}
\begin{eqnarray}
{\cal H} 
&=&
J_1 \sum_i {\bm \sigma}_i \cdot {\bm \sigma}_{i+1}
+
J_2 \sum_i {\bm \sigma}_i \cdot {\bm \sigma}_{i+2}.
\label{J1J2Hamiltonian}
\end{eqnarray}
Identifying $\Psi=\arctan(J_2/J_1)$, the optimal coherent state for this Hamiltonian is  ferromagnetic for $\arctan(-1/4)<\Psi<3 \pi/2$, antiferromagnetic for $ \arctan(1/4)>\Psi > -\pi/2$. Between $\arctan(-1/4)$ and $\arctan(1/4) $,  the optimal coherent state is a spiral that winds up from infinite pitch in the ferromagnet, to pitch given by twice the lattice constant in the antiferromanget. At the Majumdar-Ghosh point, $J_2=J_1/2$, the exact groundstate  is given by the equal superposition of the two singlet decorations of the line. These states are all captured by the simplified ansatz of (\ref{SimpleMPS}): magnetic states by $n^z=1$ and $|{\bf l} \rangle$ the appropriate spin coherent state; the Majumdar-Ghosh state by $n_1=-n_2=1/\sqrt{2}$ and $|{\bf l} \rangle$ alternating between $| {\uparrow} \rangle$ and $ |{\downarrow} \rangle$.

\begin{figure}[t!]
\begin{center}
\includegraphics[width=0.8 \linewidth]{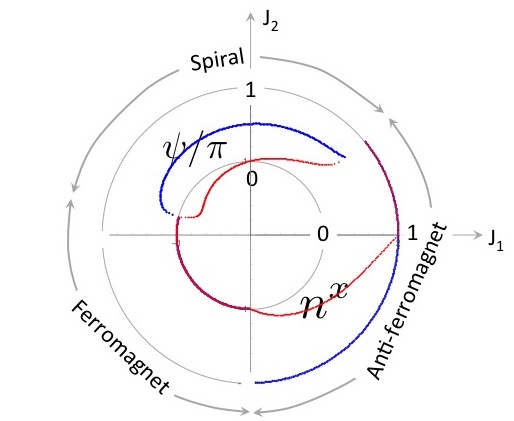}
\caption{{\it Spiral and Entanglement Structure of $J_1$-$J_2$ model:} Properties of the $J_1$-$J_2$ model plotted as a function of the angle $\arctan(J_1/J_2)/\pi$. The pitch angle $\psi$ denotes the rotation of the local spin basis from one site to the next. $n^x$ provides a measure of entanglement - $n^x=0$ corresponding to zero entanglement. The pitch follows roughly the behaviour of the optimum coherent state except that the spiral region does not extend quite as far towards low $J_1$.  There is a change in entanglement structure at the point $J_1=0$ which is driven further by the inclusion of the effects of zero-point fluctuations.}
\label{fig.J1J2Model}
\end{center}
\end{figure}

 The saddle-point of Eq.(\ref{J1J2Hamiltonian}) over the ansatz (\ref{SimpleMPS}) is given by $\lambda=1/2$, uniform ${\bf n}$, and a spiral configuration of the vectors ${\bf l}$, rotating by an angle $\psi_i$ between site $i$ and site $i+1$. Taking ${\bf l}$ to lie in the $xy$-plane and a uniform rotation, the expectation of the Hamiltonian is 
\begin{eqnarray*}
{\cal H}/N
&=&
 \frac{ J_1}{2} \left(
 ({n^z})^2+1+n^x\right) \cos \psi 
 + J_2 ({n^z})^2 \cos 2 \psi
 -\frac{J_1}{2} n^x .
\end{eqnarray*}
The optimal $n$ and $\psi$ are shown in Fig.~\ref{fig.J1J2Model}. Note the importance of the points $J_1=0$ and the Majumdar-Ghosh point, $J_1=2J_2$. 
Allowing for quantum fluctuations about the saddle-point configurations reveals a phase transition at $J_1=0$, where zero-point fluctuations favour  $\lambda$ away from $1/2$. To demonstrate this, we calculate the expectation of the Hamiltonian and zero-point fluctuations about it for an arbitrary value of $\lambda$. The result is to be considered an effective potential for $\lambda$.

The saddle-point equations with  $\lambda \ne 1/2$ lead to a staggered spiral with both ${\bf n}_i$ and $\psi_i$ alternating between two values from site to site: 
$n^x_i \rightarrow n^x_e, \; n^x_o$ and $\psi_i \rightarrow \psi_e,\; \psi_o$ on even and odd sites, respectively. 
To leading order in $X=\frac{J_1}{2J_2}$, the solutions can be expanded as
\begin{eqnarray*}
n^x_{e/o}&=&  \frac{X}{2} \left(1 + \frac{X}{2} \right) \mp X (\lambda-1/2)  \\
\delta \psi_{e/o} &=& \frac{X}{2} \left(1 + \frac{X}{4} \right) \mp (2-X) (\lambda-1/2) 
\end{eqnarray*}
The expectation of the Hamiltonian is then given by
$${\cal H}/N= J_2 \left[ -1 - \frac{3}{4} X^2 - \frac{1}{4} X^3 + X^2 (\lambda -1/2)^2 \right]. $$

{\it Beyond Mean-field Theory:} Our analysis so far is a mean-field theory over the restricted bond-order 3 ansatz of Eq.(\ref{SimpleMPS}). Even with the minimal entanglement that it embodies, qualitatively new features emerge if we consider an expansion in fluctuations about this state. This may be acheived by considering the effect of a restricted set of local unitaries $ U_{\bm m}= \exp[ i {\bm m} \cdot {\bm \sigma}/2] ,$
 with ${\bm m} \cdot {\bm l}=0$ so that ${\bm m}= m_1 {\bm \theta} + m_2 {\bm \phi}$, where 
 ${\bm \theta}$ and ${\bm \phi}$ are unit vectors perpendicular to ${\bm l}$ and proportional to $\partial_\theta {\bm l}$ and $\partial_\phi {\bm l}$, respectively [see Appendix \ref{App: Columnar VBS to Neel Transition}].
%
After transforming to the Fourier components of the fields on odd and even sites,
\begin{eqnarray*}
{\bm m}_{e,\alpha} =  \sum_{p} e^{i p \alpha} {\bm m}_{e,p} \hbox{  and  }
{\bm m}_{o,\alpha} = \sum_{p} e^{i p( \alpha+1/2)} {\bm m}_{o,p}, 
\end{eqnarray*}
the spin-wave Hamiltonian can be written as
\begin{eqnarray}
{\cal H}
&=&
\sum_k 
\left(
\begin{array}{c}
m_{1e,k} \\ m_{1o,k} \\ m_{2e,k} \\ m_{2o,k}
\end{array}
\right)^T
\left(
\begin{array}{cccc}
A & B & 0 & 0\\
\bar B & A & 0 & 0\\
0& 0& C& D\\
0& 0& \bar D & C
\end{array}
\right)
\left(
\begin{array}{c}
m_{1e,-k} \\ m_{1o,-k} \\ m_{2e,-k} \\ m_{2o,-k}
\end{array}
\right)
\nonumber\\
A& \approx &  J_2  \left[  1+ \cos k  \right]
\nonumber \\
B&\approx &  2J_2 X \cos (k/2) 
\nonumber\\
C& \approx &  J_2 \left[ 1-\cos k \right]
\nonumber\\
D& \approx & -4 J_2 X i (\lambda-1/2) \sin (k/2),
\label{J1J2spinwaveH}
\end{eqnarray}
where the approximate equalities apply to leading order in $X$ [see Appendix \ref{App: J1J2model}]. 
The Berry phase calculated in the same limit is given by the usual spin-wave Berry phase. Expressing the Hamiltonian in terms of the bosonic fields $a= a_1+i a_2 = m_{1e}+ i m_{2e} $ and $b= b_1+i b_2 =   m_{1o}+ i m_{2o} $, it can be diagonalised with a combined unitary and Bogoliubov transformation. After 
these manipulations [see Appendix \ref{App: Columnar VBS to Neel Transition}], we find that the zero point energy in limit of small $k$ , but still with $k \gg X$ is given by
\begin{equation}
 E_{zero}= J_2 \sum_k \left[\frac{k}{\sqrt{2}} -1 - \frac{8 X (\lambda-1/2)^2}{k\sqrt{2}} \right].
 \label{ZeroPointEnergy}
 \end{equation}
Notice that the zero-point energy at $\lambda=1/2$ is negative. This is the same as the case of the antiferromanget where quantum fluctuations lower the energy of the classical antiferromangetic groundstate. The $\lambda$-dependent correction to this goes as $-8(\lambda-1/2)^2 /\sqrt{2} J_1 \log |J_1/J_2|$. The important point about this is that for negative $J_1/J_2$, the potential for $\lambda$ favours $\lambda=1/2$. For positive $J_1/J_2$, the potential favours $\lambda=0$ or $1$. In fact, in the latter case, the change in entanglement structure takes us away from the regime of validity of our expansions. However, this simple calculation serves to demonstrate that zero-point fluctuations can drive a change in the entanglement structure by an order-by-disorder mechanism. Note that this behaviour could not have been found from the coherent state path integral (without an instanton expansion) nor could it have been seen from the bond order 3 MPS ansatz of Eq.(\ref{SimpleMPS}). This illustrates the power of the MPS field theory.

\subsection{Antiferromagnetic Chain}
Similar effects occur in the antiferromanget at $J_2=0$. The physics embodied in the spatially-dependent $\theta$-terms first identified by Haldane\cite{haldane19883} is already contained in the fluctuation-corrected saddle point recovering the large-$N$ results of Read and Sachdev\cite{read1989valence,read1991large}. The expectation of the antiferromagnetic Hamiltonian with the state Eq.(\ref{SimpleMPS}) is independent of $\lambda$ and optimised by an antiferromagnetic configuration of ${\bf l}$. The dependence upon ${\bf n}$ is given by
$$
{\cal H}= -J \sum_{i}  \left(  1 + ({n^z})^2 + 2 n^x \right),
$$
which is minimised by $n^z=0$, $n^x=1$ (and is extremely flat, $\propto ({n^z})^4/4 $ near $n^z$=0).   Allowing the zero-point energy of fluctuations breaks the degeneracy in $\lambda$.

The effective Hamiltonian for fluctuations in ${\bf l}$ is that of an antiferromagnet with alternating coupling strength. The resulting zero-point energy is proportional to $\sqrt{ \lambda (1 - \lambda)}  \sqrt{ ( 1+n^x )/(1-n^x)}$ near $n^z=0$\footnote{To a first approximation, the Berry phase for ${\bf l}$ is zero (since $n^z=0$). However, if we keep ${\bf n}$ general and expand about the antiferromagnetic saddle-point for ${\bf l}$, fluctuations in ${\bf l}$ change effective potential for ${\bf n}$ and $\lambda$. }. This residual dependence upon ${\bf n}$ together with the zero-point energy favours $n_x=1$ and $\lambda=0$ or $1$ - that is it picks out one of the two singlet patternings of the chain. This is just what Read and Sachdev found by including the Haldane $\theta$-terms in a large-$N$ expansion. The precise situation in our case is delicate; the coefficient of the Berry phase for ${\bf l}$ strictly goes  to zero in the limit the $n^z \rightarrow 0$. However, $n$ is strongly fluctuating and our expansion about a translationally invariant saddle point has neglected this. The essence of our result is that fluctuations have picked out an entanglement structure by an order-by-disorder mechanism.

\section{Conclusions and Outlook}
\label{sec.Conclusions}

We have extended the Feynman path integral to encompass entangled paths, {\it i.e. } sequences of matrix product states rather than the un-entangled product states of the usual Feynman path integral. By conjoining methods from tensor networks and field theory, this affords both conceptual and analytical insights. 

Amongst its conceptual advantages, the Feynman path integral demonstrates very succinctly how the classical limit emerges through the dominance of saddle-point configurations in the path integral. These saddle point configurations obey the classical dynamical equations. In the present generalisation, the equations of motion of the saddles are the quantum dynamics projected onto the variational manifold of the matrix product states. This is the time-dependent variational principle. The variational manifold can in fact be interpreted as a new emergent semi-classical phase space;\cite{PhysRevLett.107.070601,2012arXiv1210.7710H} and the variables of the matrix product state as emergent semi-classical collective coordinates of the system. Field theoretical considerations allow us to understand when these additional variables are necessary to describe the system.

The central notion is that of adiabatic continuity. Familiar in the context of Fermi liquid theory - where the Fermi liquid is adiabatically continuous with the Fermi gas - a low-bond-order saddle point is said to be adiabatically continuous with the full saddle point if the latter may be obtained by a perturbative (though possibly large) dressing of the low bond order saddle. Remarkably, even in situations in which entanglement diverges - such as at a quantum critical point - we can often use a product state field theory to good effect. The critical state is adiabatically connected to a state without any entanglement!\footnote{I include formally non-perturbative ways of controlling the interactions, such as $\epsilon$- and $1/N$-expansions as perturbative corrections here.} 

This is not the case for unconventional phase transitions, such as deconfined quantum critical points. The critical state is not adiabatically connected to a product state and this is intimately related to the inability to describe the transition with a conventional Ginzburg-Landau theory. Typically, such systems are described using emergent gauge fields. We have shown that a path integral over a low bond-order MPS provides an alternative (in one dimension). The MPS tensors become an effective Ginzburg-Landau order parameter for the transition and the path integral is dual to the emergent gauge description. The spirit of the MPS path integral is different from the conventional manner in which MPS are used; usually bond-order is increased to obtain numerical convergence. Here, the bond-order is increased until the mean-field saddle-point is adiabatically connected to the full saddle-point. We have illustrated this with a simple model of a valence-bond solid to Ne\'el transition. We used a simple restriction of bond-order three MPS and showed that the transition is explicitly a transition in entanglement structure and that, moreover, the critical theory of the transition is approximately captured by a transverse field Ising model with easy axis anisotropy. 

An intriguing feature of the MPS path integral is its ability to capture in its saddle point (and possibly perturbative fluctuations about it) physical features that can only be found in the conventional field theory by a proliferation of instantons. Being a (restricted) sum of product states, the MPS can describe tunneling trajectories in real time. We illustrated this in the one-dimensional $J_1$-$J_2$-model where a transition in the mean-field spiral phase - driven in the product state field theory by a proliferation of instantons - was revealed to be a change in entanglement structure driven through an order-by-disorder effect of spin-wave excitations. 

The results presented here scratch the surface of what might be achieved with such path integrals over entangled states. For example, in Section \ref{Sec:J1J2Model}, we calculated fluctuation corrections to an analytically determined saddle-point over a restricted class of bond-order three MPS. One may also envisage using this same approach to improve low-bond order numerically optimised MPS states, just as fluctuation corrections are used to improve product state/mean field approximations\footnote{Albeit, the correction is usually applied to the Hamiltonian rather than the state}. As emphasised above, an MPS is a restricted sum of product states. As bond-order is increased, the number of product states that are summed over is increased and the approximation converges, since any point in Hilbert space may be constructed as a sum of product states. At low-bond order, the product states added together in this  way will be substantially different. At some bond-order (say $D^*$), however, the additional product states added are only small(-ish) deformations of those already included. This is the point at which the MPS saddle point is adiabatically continuous with the actual saddle point. Increasing bond-order is a relatively cumbersome way of allowing for the fact that the saddle is captured by a sum of states that deviate only slightly from a point on the variational manifold of bond-dimension $D^*$. A field theoretical expansion in fluctuations [viewed either as a correction to the state or to the Hamiltonian acting on the variational manifold] may, in some circumstances, be a more convenient way of proceeding beyond the bond dimension $D^*$\footnote{We will communicate results for the transverse field Ising model in a forthcoming work.}.

Another potential application of the machinery of MPS path integrals is to the study of open quantum systems. It is widely understood that coupling to the environment degrades and constrains the amount of entanglement that is usefully preserved in a quantum system\cite{crowley2014quantum}. Constructing a Keldysh field theory over MPS states has the potential to reveal how this proceeds through the effects of the environment on quantum trajectories.

Beyond these immediate applications, there are some more challenging directions in which the MPS path integral warrants development. Foremost amongst these are the renormalization of the MPS path integral and its extension to higher dimensions. The path integrals that we have constructed in this work, though having an unusual origin, are
rather conventional in structure and ought to be amenable to taking continuum limits and renormalization in the usual way. The interpretation of such a scheme is interesting; many of the fields that are to be renormalized describe entanglement structure. How the results of such an analysis relate to the many insights on the renormalization of tensor networks\cite{verstraete2008matrix,Xie:2009rk,xie2012coarse,qi2013exact,2009PhRvA..79d0301P} remains to be determined, as does the relationship to the quantum renormalization group\cite{Lee:2013im}.

Extension of the entangled path integral to higher dimensions will open up a broader range of potential applications. In the case of the MPS field integral,
the finite amount of data required to parametrize the environment of a local operator allowed us to construct a spatially local field integral. The most natural extension of MPS states to higher dimensions, projected entangled pair states (PEPS), require an infinite amount of data to describe the environment in the thermodynamic limit. This prevents the construction of a spatially local path integral\footnote{The usual numerical approach is to introduce additional refinement parameters ---  for example truncating the bond-order of an MPS\cite{jordan2008classical} or corner transfer matrix\cite{orus2009simulation} description of the environment. This truncation effectively projects the PEPS states to a more restricted sub-manifold\cite{jordan2008classical,orus2009simulation}. This implicit identification of the variational manifold makes it much trickier to construct  a higher dimensional field integral.}. Luckily there are potentially several ways to circumvent this difficulty. 
Firstly, explicit restrictions of PEPS akin to that imposed on MPS in Sec. \ref{sec:ParamandExamples} may render the PEPS efficiently contractible; the restricted MPS of Eq.(\ref{SimpleMPS}) has a natural extension to higher dimensions. 
An alternative approach is to construct the path integral over a set of states described by finite depth circuits. Such states are by design local and finitely contractable in any dimension, although this is at the expense of breaking explicit translational invariance in the description. Finding an appropriate measure is straightforward; it is the Haar measure over each of the unitary elements of the circuit. A field theoretical renormalisation of this path integral will bear interesting relation to a path integral over hierarchical networks such as the multiscale entanglement renormalization ansatz (MERA)\cite{2007PhRvL..99v0405V} . MERA is also efficiently contractible by design and so leads naturally to spatially local path integrals. Identifying tractable problems for the use of such path integrals is a fascinating direction for future study. 

Moreover, a path integral constructed over MERA points towards some longer-term goals.  One of the advantages of the Feynman path integral is the way in which it emphasises the importance of classical configurations. Perhaps ultimately a Feynman path integral over entangled paths can motivate the emergence of a hierarchical network --- and the geometry that it implies\cite{Swingle:2012nj} --- as semi-classical collective coordinates of many-body quantum systems.  
This would help formalise several suggestive links: between the ideas of tensor networks and AdS/CFT\cite{Swingle:2012nj}, between conventional renormalization and entanglement based schemes\cite{verstraete2008matrix,Xie:2009rk,xie2012coarse,qi2013exact,2009PhRvA..79d0301P}, and the quantum renormalisation group\cite{Lee:2013im,lee2010holographic,lee2011holographic}. 

Theoretical physics has benefited of late from a remarkable confluence of ideas. Diverse sub-disciplines from quantum information and condensed matter, to string theory and gravitation have been addressing overlapping questions. A new language is evolving to express the resulting insights. Here we have combined ideas from field theory with those from tensor networks, melding the facility of field theory in understanding the emergence of classicality with that of tensor networks in encoding subtle aspects of classicality in quantum systems. The result has the potential to shed light upon a  range of problems.

\appendix
\section{Parametrising the MPS manifold} \label{App: Parametrising MPS states}
In Ref.(\onlinecite{wouters2013thouless}), Wouters {\it et al.} provide a compact parametrization of the entire MPS manifold. In essence, these authors use the Thouless theorem to exponentiate Eq.(\ref{MPStangentvec}). In this Appendix, we demonstrate the equivalence of Eq.(\ref{SpinorMPS}) to the expression provided in Ref.(\onlinecite{wouters2013thouless}). Using a Dirac notation for the spin indices and a matrix notation for the auxiliary indices,
$|A_{ab} \rangle =\sum_\sigma A^\sigma_{ab} |\sigma \rangle$, 
$\langle A_{ab} | =\sum_\sigma {A^\sigma_{ab}}^\dagger  \langle \sigma |.$
the general parametrization of the MPS tensor on a particular site is given by
\begin{equation}
|A(x) \rangle = \exp \big[ |dA(x)\rangle \langle A| - |A \rangle \langle dA(x)| \big] |A \rangle,
\label{Wouters}
\end{equation}
where site and auxilliary indices have been suppressed for clarity - the definitions of $x$ and $dA$ are given in the main text. The relationship of this to Eq.(\ref{SpinorMPS}) is readily demonstrated by expanding Eq.(\ref{Wouters}) using the orthogonality of $|dA\rangle$ and $|A\rangle$ ($\langle dA| A \rangle=0$) and the normalization of the reference MPS ($\langle A| A \rangle={\bm 1}$, {\it i.e. } ${A^\sigma_{ba}}^* A^\sigma_{bc} = \delta_{ac}$);
\begin{eqnarray*}
|A(x) \rangle
&=& 
|A \rangle \cos {\bm M}
+
 |dA(x) \rangle {\bm M}^{-1} \sin {\bm M},
 \nonumber\\
\hbox{ with }
& &
{\bm M} =\sqrt{  \Lambda^{-1/2}  {x^{\sigma\ne1}}^\dagger x^{\sigma \ne 1}  \Lambda^{-1/2}}
\end{eqnarray*}
a matrix in auxiliary space.
Finally, by writing $dA$ in terms of the MPO, Eq.(\ref{MPStangentvec}), we identify 
\begin{eqnarray*}
z^1_{ab} 
&=&
\cos(  {\bm M})_{ac} \Lambda^{1/2}_{cb}
\nonumber\\
z^{\delta \ne 1}_{ab}
&=&
x^{\delta\ne 1}_{ac} \Lambda^{-1/2}_{cd} 
\left( {\bm M}^{-1} \sin {\bm M} \right)_{de}
 \Lambda^{1/2}_{eb},
\end{eqnarray*}
which automatically satisfy the constraints in Eq.(\ref{SpinorMPS}).

The simplest example of this parametrisation is that found for spin 1/2 when taking the reference MPO ${\cal A}^{\sigma, \delta}_{a,b}= \delta_{\sigma,\delta} \delta_{a,b}$. In this case, the spinor representation of the MPS reduces to a simple rescaling by the environment tensor;
$
A^{\sigma}_{ab}(z)= z^\sigma_{ac} \Lambda^{-1/2}_{cb}.
$
The representation of Wouters {\it et al.} can be written in this case as
$$
|A(x)\rangle
=
\exp \left[ x^\downarrow \Lambda^{-1/2} \hat \sigma^+ - \Lambda^{-1/2} {x^\downarrow}^\dagger \hat \sigma^- \right]
|\uparrow \rangle,
$$
where we have explicitly written the spin structure in Dirac notation and used an implicit matrix structure in auxiliary space with indices suppressed for clarity.

As emphasised in Section \ref{sec:Geometry}, the MPS (\ref{}) can be considered a subset of the components of a time-dependent unitary matrix $A^\sigma_{ab}= {\cal A}^{\sigma, \delta=1}_{ab}$. Again using a mixed notation in which the spin structure is indicated by Dirac notation and the auxilliary structure through an explicit matrix form, we may write 
$| A(z)_{ab} \rangle =  \hat A_{ab}(z) |  \uparrow \rangle$, with
\begin{eqnarray}
\hat A_i(z)
&=&
\left(
\begin{array}{ccc}
0 & | {\bf l}_i \rangle \langle \uparrow | & | - {\bf l}_i \rangle \langle \uparrow| \\
n_{1,i-1}| {\bf l}_i \rangle \langle \uparrow |& 0 &0 \\
n_{2,i-1}| - {\bf l}_i \rangle \langle \uparrow | & 0 & 0 
\end{array}
\right)
\nonumber\\
& &
+
\left(
\begin{array}{ccc}
0 & 0 & 0 \\
 \tilde n_{1,i-1}| {\bf l}_i \rangle \langle \downarrow |& 0 & | - {\bf l}_i \rangle \langle \downarrow | \\
 \tilde n_{2,i-1}| -{\bf l}_i \rangle \langle \downarrow | & |{\bf l}_i \rangle \langle \downarrow | & 0 
\end{array}
\right),
\nonumber\\
\label{ansatzMPO}
\end{eqnarray}
where $z \equiv z(n,{\bf l})$ and $\tilde n =(\tilde n_1,\tilde n_2)$ is an $SU(2)$ spinor that is orthogonal to the spinor $n=(n_1,n_2)$.  The mixed notation emphasises both the factorisation between auxiliary and spin structure in the simple ansatz Eq.(\ref{SimpleMPS})and also the relative sparsity of the auxiliary structure.

Finally, there are several ways of determining the Berry phase for the simple ansatz of Eq.(\ref{SimpleMPS}). One way --- as discussed in the main body of the text --- is to explicitly identify the tensors $z$ in terms of $n$ and ${\bf l}$. An alternative is to use Eq.(\ref{SimpleMPS}) directly and to evaluate the Berry phase from Eq.(\ref{ansatzMPO}) using the form
$
{\cal S}_B= \sum_i \int dt \sum_\sigma Tr \left[  \dot A_i^\sigma \Lambda_i {A_i^\sigma}^\dagger \right]
$.

\section{Manipulating Simple MPS}
\label{App: Manipulating Simple MPS}
In order to help manipulate the simple MPS state that we define in Eq.(\ref{SimpleMPS}), we note a couple of useful relations. i. We use the gauge
\begin{eqnarray*}
|{\bf l} \rangle 
&= &
\cos \theta_2 e^{-i \phi/2} |\uparrow \rangle
+
\sin \theta_2 e^{i \phi/2} |\downarrow \rangle,
\\
|-{\bf l} \rangle 
&= &
- \sin \theta_2 e^{-i \phi/2} |\uparrow \rangle
+ \cos \theta_2 e^{i \phi/2} |\downarrow \rangle.
\end{eqnarray*}
In this gauge,
$ 
\langle {\bf l}| {\bm \sigma} | - {\bf l} \rangle 
=
{\bm \theta}_{\bf l} - i {\bm \phi}_{\bf l},
$
where $\hat \theta$ and $ \hat \phi$ are unit vectors in the directions of $\partial_\theta {\bf l}$ and $\partial_\phi {\bf l}$ respectively;
\begin{eqnarray*}
{\bm \theta}
&=&
(\cos \theta \cos \phi, \cos \theta \sin \phi , \sin \theta),
\\
{\bm \phi}
&=&
(- \sin \phi, \cos \phi, 0 ).
\end{eqnarray*}

\noindent
ii. The following expectation appears in the various Hamiltonians that we consider:
\begin{eqnarray*}
 \langle {\bf l},{\bf m} | 
  {\bm \sigma}.{\bm \sigma} 
  |  - {\bf l}, - {\bf m} \rangle 
=
  {\bm \theta}_{\bf l}. {\bm \theta}_{\bf m}- {\bm \phi}_{\bf l}. {\bm \phi}_{\bf m}-i(  {\bm \theta}_{\bf l}. {\bm \phi}_{\bf m} + {\bm \phi}_{\bf l}. {\bm \theta}_{\bf m} ).
   \label{UsefulExpectation}
\end{eqnarray*}

\section{Columnar VBS to Neel Transition} \label{App: Columnar VBS to Neel Transition}
The expectation of the VBS-Ne\'el Hamiltonian  with the ansatz state Eq.(\ref{SimpleMPS}) is given by
\begin{eqnarray*}
{\cal H}
&=&
J_A \sum_{ij} 
 \left(  
 \lambda n^z_{2i-2,j} n^z_{2i,j}  
 +(1-\lambda)   
 \right) {\bf l}_{2i-1,j} \cdot {\bf l}_{2i,j}  
  \\
 & &
- J_A \sum_{ij} (1-\lambda)
\frac{  n_{2i-1,j}^x+i n_{2i-1,j}^y }{2}
\\
& &
\;\;\;\;\;\;\;\;\;\;\;
\times
  \langle {\bf l}_{2i-1,j} , {\bf l}_{2i,j} | 
  {\bm \sigma}\cdot {\bm \sigma} 
  |  - {\bf l}_{2i-1,j}, - {\bf l}_{2i,j} \rangle 
+ c.c.
\\
 & &
 + J_B \sum_{ij} 
 \left( 
  (1-\lambda) n^z_{2i-2,j} n^z_{2i,j}    +  \lambda 
  \right) 
  {\bf l}_{2i,j} \cdot {\bf l}_{2i+1,j}
 \\
 & &
 - J_B \sum_{ij} \lambda
 \frac{ n_{2i}^x+i n_{2i}^y }{2}
 \\
& &
\;\;\;\;\;\;\;\;\;\;\;
\times
  \langle {\bf l}_{2i,j},{\bf l}_{2i+1,j} | 
  {\bm \sigma} \cdot {\bm \sigma} 
  |  - {\bf l}_{2i,j},- {\bf l}_{2i+1,j}  \rangle 
   + c.c
   \\
& &
+ J_A \sum_{ij} 
 \left(  
 (1-\lambda) {n^z_{2i+1,j}}  + \lambda n^z_{2i,j} 
 \right)
 \\
 & &
 \;\;\;\;\;\;
 \times  \left(  
 (1-\lambda) {n^z_{2i+1,j+1}} + \lambda n^z_{2i,j} 
 \right) 
 {\bf l}_{2i+1,j} \cdot {\bf l}_{2i+1,j+1} 
 \\
 & &
+J_A \sum_{ij}
 \left(  
 \lambda {n^z_{2i,j}} +(1- \lambda) n^z_{2i-1,j} 
 \right)
 \\
 & &
 \;\;\;\;\;\;\;\;\;\;
 \times
 \left(  
 \lambda {n^z_{2i,j+1}}  +(1- \lambda) n^z_{2i-1,j+1} 
 \right)
 {\bf l}_{2i,j} \cdot {\bf l}_{2i,j+1}
\end{eqnarray*}
The saddle point of this Hamiltonian is given in the main body of the text. 

\vspace{0.1in}
\noindent
 {\it Fluctuation Corrections} in the approximation where ${\bf l}$ is fixed and ${\bf n}$ allowed to vary are given in the main text. In the alternative approximation, where ${\bf n}$ is fixed and uniform, ${\bf l}$ is allowed to vary, and taking $\lambda=1$, the Hamiltonian expectation reduces to 
 \begin{eqnarray*}
 {\cal H} 
 &=&
 J_A \sum_{i'j}
 {n^z}^2 {\bm l}_{2i'-1,j} \cdot {\bm l}_{2i',j}
 + J_A \sum_{ij}
 {n^z}^2 {\bm l}_{i,j} \cdot {\bm l}_{i,j+1}
 \\
 & &
 +
 J_B \sum_{i'j}
  {\bm l}_{2i',j} \cdot {\bm l}_{2i'+1,j}
 \\
 & &
 -
  J_B \sum_{i'j}
 {n^x} ( {\bm \theta}_{2i',j} \cdot {\bm \theta}_{2i'+1,j} - {\bm \phi}_{2i',j} \cdot{\bm \phi}_{2i'+1,j} )
 \end{eqnarray*}
with a Berry phase given by
$
{\cal S}_B 
=
\sum_i
\int dt 
n^z \langle {\bf l}_i | \dot {\bf l}_i \rangle.
$
A Ne\'el transformation [which changes the sign of all terms and replaces the relative minus sign in the final term  by a plus signs ] and the spin-wave expansion
\begin{eqnarray}
{\bm l} 
&\rightarrow& 
{\bm l}' \approx {\bm l} \left(1-{\bm m}^2 /2\right) - {\bm m} \times {\bm l},
\nonumber\\
{\bm \theta} 
&\rightarrow& 
{\bm \theta}' \approx {\bm \theta} \left(1-m_2^2 /2\right) +m_2 {\bm l} + m_1 m_2 {\bm \phi}/2,
\nonumber\\
{\bm \phi} 
&\rightarrow& 
{\bm \phi}' \approx {\bm \phi} \left(1-m_1^2/2 \right) -m_1 {\bm l} + m_1 m_2 {\bm \theta}/2,
\label{SpinwaveExpansion}
\end{eqnarray}
[which follow from the action of the unitary operator
$ U_{\bm m}= \exp[ i {\bm m} \cdot {\bm \sigma}/2] $
with ${\bm m} \cdot {\bm l}=0$; ${\bm m}= m_1 {\bm \theta} + m_2 {\bm \phi}$] 
results in the spin-wave Hamiltonian
\begin{eqnarray*}
{\cal H}
&=&
J_A {n^z}^2 \sum_{i'j} ({\bm m}_{2i'-1,j} - {\bm m}_{2i',j})^2/2
\\
& &
+
J_A {n^z}^2 \sum_{ij} ({\bm m}_{i,j+1} - {\bm m}_{i,j})^2/2
\\
& &
+
J_B (1+{n^x}) \sum_{i'j} ({\bm m}_{2i',j} - {\bm m}_{2i'+1,j})^2/2.
\end{eqnarray*}
This is basically as one would obtain for a two-dimensional antiferromanget, except that the coupling constants along the rungs alternate between $J_A {n^z}^2$ and  $J_B (1+n^x) $.
Long wavelength fluctuations average this to $-[J_A {n^z}^2+J_B (1+n^x)]/2 $.
Finally, using the Haldane map and taking a continuum limit yields 
\begin{eqnarray*}
{\cal S} 
&=&
\frac{1}{2}  \int dt d {\bf x} \left[ \chi_0|\partial_t {\bf m}|^2 
+\rho_x (\nabla_x{\bf m})^2 +\rho_y(\nabla_y{\bf m})^2 
\right]
\\
\hbox{ with } & & \rho_x= [J_A {n^z}^2+J_B (1+n_\perp)]/2 
\hbox{ and } \rho_y=J_A {n^z}^2,
\\
\hbox{ and } & &\chi_0= \left[ 2a^2 [3J_A {n^z}^2+J_B (1+n_\perp)] \right]^{-1},
\end{eqnarray*}
from which one obtains zero-point correction to ${\bf l}$ as
\begin{eqnarray*}
\langle |{\bf l}|^2 \rangle
&=&
1- \frac{\Lambda}{2 \pi} \frac{1}{ \sqrt{ \chi_0 \sqrt{\rho_x \rho_y}}}.
\end{eqnarray*}
These corrections diverge as the VBS-Ne\'el transition is approached, since
$n^z \rightarrow 0$ and $\rho_y \rightarrow 0$. Self-consistent treatment of these fluctuations alongside those in ${\bm n}$ likely leads to a significant renormalisation of the critical theory presented in the main text.

\section{$J_1$-$J_2$ model} \label{App: J1J2model}
The expectation of the J1-J2 Hamiltonian, Eq.(\ref{J1J2Hamiltonian}), with the ansatz Eq.(\ref{SimpleMPS}) is given by
\begin{eqnarray*}
{\cal H}
&=&
 J_1 \sum_i
 \left( \lambda_{i-1} n^z_{i-1} n^z_{i+1} +\lambda_{i} \right)  {\bm l}^i \cdot {\bm l}^{i+1} 
 \\
 &+&
  J_1 \sum_i
 \lambda_i \frac{n^x_i +i n^y_i}{2} ({\bm \theta}_i+i {\bm \phi}_i) \cdot ({\bm \theta}_{i+1}+i {\bm \phi}_{i+1})
 +c.c.
\\
&+&
J_2 \sum_i \left(
\lambda_in^z_i  n^z_{i+2} 
 + 
\lambda_{i-1} n^z_{i-1}  n^z_{i+1} \right) {\bm l}_i \cdot {\bm l }_{i+2}.
\end{eqnarray*}
The bare saddle points of this Hamiltonian have $\lambda=1/2$ and consist of a uniform ${\bf n}$ and a spiral configuration of ${\bm l}$. It is argued in the main text that allowing for zero-point fluctuations can favour a value of $\lambda \ne 1/2$ and a consequent staggering of ${\bm n}$ and of the spiral in ${\bm l}$. 
The saddle point equations for an arbitrary $\lambda$, labelling 
$n_i \rightarrow n_o, \; n_e$, 
$\theta_i \rightarrow \theta_o,\; \theta_e$
on odd and even sites, reduce to 
\begin{eqnarray*}
\partial_{\theta_e} {\cal H}/N
&=&
-\frac{J_1}{2} (\lambda_e + \lambda_o {n_o^z}^2 ) \sin \theta_e 
+ \frac{J_1}{2} \lambda_e n_e^x \sin \theta_e
\\
& &
-J_2 (\lambda_o {n_o^z}^2 + \lambda_e {n_e^z}^2 ) \sin (\theta_o+ \theta_e) =0,
\\
\partial_{n^x_e} {\cal H}/N
&=&
-J_1 \lambda_e \left( n^x_e \cos \theta_o + \frac{1}{2} (\cos \theta_e -1) \right)
\\
& &
-2 J_2 \lambda_e n^x_e \cos(\theta_o + \theta_e)=0,
\end{eqnarray*}
with similar equations replacing $e \longleftrightarrow o$.
Noting that $\lambda_e=\lambda$ and $\lambda_o=1-\lambda$, there is also an equation obtained by optimising $\lambda$. This is solved by $\lambda=1/2$. We will show that quantum fluctuations can favour a different saddle point and so keep $\lambda$ as a free parameter for now. We interpret the result as an effective potential for $\lambda$.

We focus upon the region of the phase diagram near $J_1=0$. Around this $\psi= \pi/2+ \delta \psi$, $\lambda=1/2+\mu$, with  $\delta \psi$, $\mu$, $J_1/2 J_2=X$ all assumed small. The saddle point equations can then be reduced to 
\begin{eqnarray*}
\partial_{\theta_e} {\cal H}/N
&=&
-\frac{J_1}{2} (\lambda_e + \lambda_o {n_o^z}^2 )
+ \frac{J_1}{2} \lambda_e n_e^x 
\\
& &
+J_2 (\lambda_o {n_o^z}^2 + \lambda_e {n_e^z}^2 ) (\delta \theta_o+ \delta \theta_e) =0,
\\
\partial_{n^x_e} {\cal H}/N
&=&
+J_1 \lambda_e \left( n^x_e \delta \theta_o + \frac{1}{2} (1+\delta \theta_e ) \right)
\\
& &
+2 J_2 \lambda_e n^x_e \left(1-(\delta \theta_o + \delta \theta_e)^2/2 \right)=0,
\\
& & 
e \longleftrightarrow o
\end{eqnarray*}
whose solutions is as given in the main body of the text. 

{\it The Spin-wave Expansion} of this Hamiltonian is acheived as in Appendix \ref{App: Columnar VBS to Neel Transition} through the action of local unitaries $ U_{\bm m}= \exp[ i {\bm m} \cdot {\bm \sigma}/2] $
with ${\bm m} \cdot {\bm l}=0$ and ${\bm m}= m_1 {\bm \theta} + m_2 {\bm \phi}$.
An explicit unitary transformation of ${\bm \sigma}$ to quadratic order in ${\bm m}$ gives
$$
U^\dagger {\bm \sigma} U
=
 {\bm \sigma} (1-{\bm m}^2/2) - {\bm m} \times {\bm \sigma} + {\bm m} ({\bm \sigma }\cdot {\bm m} )/2.
$$
Using $\langle  {\bm l} |{\bm \sigma} |- {\bm l} \rangle= {\bm \theta} -i {\bm \phi} $ one finds after a bit of re-arranging that 
$\langle {\bm l} |U^\dagger {\bm \sigma} U | -{\bm l} \rangle ={\bm \theta}' -i {\bm \phi}' $ with 
${\bm \theta}'$ and ${\bm \phi}'$ given by Eq.(\ref{SpinwaveExpansion}).

We can use these results to perform a spin-wave expansion about the staggered spiral state;
\begin{eqnarray*}
 {\cal H}
&=&
 \sum_i [\tilde J_{1,i}- K_{i}]
   \cos \theta_i 
 +  \sum_i K_{i}
  (  \cos \theta_i -1)
 \\
 & &
 +
 \tilde J_2 \sum_i 
 \cos( \theta_{i+1}+\theta_i ) 
 \\
  & &
 - \sum_i \tilde J_{1,i}
\cos \theta_i    \left(m_{1,i}^2 +m_{1,i+1}^2
 \right)/2
 \\
 & &
 +  \sum_i
\tilde J_{1,i}  m_{1,i}m_{1,i+1}
\\
& &
 +  \sum_i
 K_{i}
(1+\cos \theta_i ) ( m_{1,i}-m_{1,i+1})^2/2
\\
& &
-
\tilde J_2 \sum_i 
 \cos( \theta_{i+1}+\theta_i )  \left(m_{1,i}^2 +m_{1,i+2}^2   \right)/2
 \\
& &
+
\tilde J_2 \sum_i 
m_{1,i}m_{1,i+1}
 \\ 
 & &
 - \sum_i
 \tilde J_{1,i}  
 \cos \theta_i  \left(m_{2,i}-m_{2,i+1}  \right)^2/2
 \\
 & &
- 
\tilde J_2 \sum_i 
 \cos( \theta_{i+1}+\theta_i )  \left(m_{2,i}-m_{2,i+2}  \right)^2/2,
 \end{eqnarray*}
 where we have defined
 \begin{eqnarray*}
\tilde J_2
&=&
 J_2(\lambda_e {n_e^z}^2+\lambda_o {n_0^z}^2),
\\
\tilde J_{1,i}
&=&
J_1 \left( \lambda_{i-1} n^z_{i-1} n^z_{i+1} +\lambda_{i}(1  +   n^x_{i} )  \right),
\\
K_{i}
&=&
J_1  \lambda_{i} n^x_{i} .
\end{eqnarray*}
After Fourier transforming, this can be reduced to
\begin{eqnarray*}
{\cal H}
&=&
\sum_k 
\left( m_{1e,k},m_{1o,k},m_{2e,k},m_{2o,k} \right)
\\
& &
\;\;\;\;\;\;\;\;
\times
\left(
\begin{array}{cccc}
A & B & 0 & 0\\
\bar B & A & 0 & 0\\
0& 0& C& D\\
0& 0& \bar D & C
\end{array}
\right)
\left(
\begin{array}{c}
m_{1e,-k} \\ m_{1o,-k} \\ m_{2e,-k} \\ m_{2o,-k}
\end{array}
\right)
\end{eqnarray*}
with 
\begin{eqnarray*}
\\
A&=&
 - \frac{1}{2}
 \left[  \tilde J_{1e\theta} +\tilde J_{1o\theta}  - K (\cos \theta_e + \cos \theta_o) \right]
 \\
 & &
 \;\;\;\;\;\;\;
 +K -  \tilde J_{2 \theta}  + \tilde J_2 \cos k,
 \\
B&=&
 \tilde J_{1} \cos(k/2)
 \\
 & &
 - \frac{1}{2} K \left[(1+\cos \theta_e )e^{-ik/2}+(1+\cos \theta_o )e^{ik/2} \right] ,
\\
C&=&
 - \frac{1}{2}  \left[   \tilde J_{1e\theta} +\tilde J_{1o\theta} + 2 \tilde J_{2 \theta} - 2 \tilde J_{2\theta} \cos k  \right],
\\
D&=&
\frac{1}{2}   \left[  \tilde J_{1e\theta} e^{-ik/2}+\tilde J_{1o\theta} e^{ik/2} \right],
\end{eqnarray*}
and where
\begin{eqnarray*}
 \tilde J_{1e\theta}=\tilde J_{1} \cos \theta_e,\;\;\;
  \tilde J_{1o\theta}=\tilde J_{1} \cos \theta_o,\;\;\;
\tilde J_{2\theta} = \tilde J_2  \cos( \theta_{e}+\theta_o ).
\end{eqnarray*}
Expanding near to $J_1=0$ we find to order $X^2$ that 
 \begin{eqnarray*}
 \tilde J_2 
 &=&
 J_2 \left[1-\lambda(1-\lambda) X^2    \right]
 \\
 K_e
 &=& K_o=
2  J_2 X^2 \lambda(1-\lambda)
 \\
 \tilde J_{1,e} &=& 
 \tilde J_{1,o}=2 J_2  X \left[1+X\lambda(1-\lambda)\right],
 \end{eqnarray*}
 where the equality between $J_{1,e}$ and $J_{1,o}$ is true to order $\mu X^5$.
Using these expansions [and $\delta \psi_{e/o}=X/2 \pm (\lambda-1/2)(2-X)$], we obtain
\begin{eqnarray*}
A
& \approx & 
J_2 
\left[ 
1+ \cos k - X^2(5+ \cos k )/4
\right],
\\
B
&\approx & 
J_2
2X \left[
\cos (k/2 +i  X (\lambda-1/2) \sin (k/2)/2)
\right],
\\
C
& \approx &
 -J_2 
\left[
\cos k-1 - X^2
\right],
\\
D
& \approx & 
J_2 
X \left[ X \cos (k/2) - 4i (\lambda-1/2) \sin (k/2) \right].
\end{eqnarray*}
These reduce to the forms given in Eq.(\ref{J1J2spinwaveH}) to leading order in $X$. 
Notice that the only linear dependence upon $\lambda$ comes from the terms $B$ and $D$ - and indeed only from the latter to linear order in $X$.

{\it The Berry Phase} in the same limit is given by
\begin{eqnarray*}
{\cal S} 
&=&
\int dt \sum_i \left( \lambda_{i-1} n_{i-1}^z + \lambda_i n^z_i \right) 
\langle {\bm l}_i | \dot {\bm l}_i \rangle
\\
&=&
\underbrace{
 \left[1 - \frac{1}{2} X^2\left(\frac{1}{4} -\mu^2 \right) \right]
 }_{\#}
 \int dt \sum_i
 \langle {\bm l}_i | \dot {\bm l}_i \rangle .
\end{eqnarray*}
In principle, the normalization $\#$ modifies the relationship between the spin-wave vector ${\bm m}$ and the bosonic fields used to diagonalize the spin-wave Hamiltonian:
\begin{eqnarray*}
a&=& a_1+i a_2 = \sqrt{\#} ( m_{1e}+ i m_{2e} ),
\\
b&=& b_1+i b_2 = \sqrt{ \#} ( m_{1o}+ i m_{2o} ).
\end{eqnarray*}
However, near to $J_1=0$, $\#=1+O(X^2)$ and the bosonic mapping reduces to the usual form used in the main text. Expressed in terms of the bosonic fields $a$ and $b$, the Hamiltonian takes the form 
\begin{eqnarray*}
{\cal H}
=
\sum_k 
\left( a^\dagger_k,b^\dagger_k,a_{-k},b_{-k} \right)
\left(
\begin{array}{cc}
{\bm \alpha} & {\bm \beta}\\
\bar {\bm \beta}& {\bm \alpha} 
\end{array}
\right)
\left(
\begin{array}{c}
a_k \\ b_k \\ a^\dagger_{-k} \\ b^\dagger_{-k}
\end{array}
\right)
-
\sum_k Tr {\bm \alpha}
\end{eqnarray*}
with
\begin{eqnarray*}
{\bm \alpha}
&=&
\frac{1}{4 \#}
\left(
\begin{array}{cc}
A+C & B+D \\
\bar B + \bar D & A+C 
\end{array}
\right),
\\
{\bm \beta}
&=&
\frac{1}{4 \#}
\left(
\begin{array}{cc}
 A-C& B-D\\
 \bar B - \bar D & A-C \\
\end{array}
\right).
\end{eqnarray*}

{\it Diagonalising:} The bosonic Hamiltonian  can be diagonalised with a combined unitary and Bogoliubov transformation of the form
\begin{eqnarray*}
& &
\left(
\begin{array}{c}
\alpha_k \\ \beta_k \\ \alpha^\dagger_{-k} \\ \beta^\dagger_{-k}
\end{array}
\right)
=
\left(
\begin{array}{cc}
M_1 & M_2\\
M_2& M_1
\end{array}
\right)
\left(
\begin{array}{c}
a_k \\ b_k \\ a^\dagger_{-k} \\ b^\dagger_{-k}
\end{array}
\right)
\end{eqnarray*}
In order to preserve the bosonic commutation relations, we require
$$
M_1 M_1^\dagger - M_2 M_2^\dagger = {\bm 1},
\;\;\;\;
M_2 M_1^\dagger=M_1 M_2^\dagger.
$$
These conditions are solved by the choice
\begin{eqnarray*}
& & M_1 = W U Z^\dagger, 
\;\;
M_2 = W V Z^\dagger 
\end{eqnarray*}
with $ W W^\dagger=Z Z^\dagger = {\bm 1}$, $U$ and $V$ diagonal, and $U^2-V^2={\bm 1}$.
Requiring that the Hamiltonian is diagonalised gives the following condition and diagonalized Hamiltonian:
\begin{eqnarray*}
U \tilde {\bm \alpha} V+ V \tilde {\bm \alpha} U= V \tilde {\bm \beta} V + U \tilde {\bm \beta} U 
\\
\epsilon 
=
W 
\left( \begin{array}{cc} 
U \tilde {\bm \alpha} U & - V \tilde {\bm \beta} U \\ 
-U \tilde {\bm \beta} V & V \tilde {\bm \alpha} V
\end{array}
\right)
W^\dagger
\end{eqnarray*}
with $\tilde {\bm \alpha} = Z^\dagger \alpha Z, \;\;\; \tilde {\bm \beta} = Z^\dagger \beta Z$. These conditions are equivalent to $\tilde {\bm \alpha} = U \tilde \epsilon U + V \tilde \epsilon V $ and
$\tilde {\bm \beta} = U \tilde \epsilon V + V \tilde \epsilon U$.
%
%
In order to find the eigen-energies and zero-point energy, it is convenient to note that
\begin{eqnarray*}
Tr[ \epsilon^2 ] 
&=&
\epsilon_1^2+\epsilon_2^2= Tr[ \tilde {\bm \alpha}^2 - {\bm \beta}^2 ] 
=8AC,
\\
det [ \epsilon^2]
&=&
\epsilon^2_1 \epsilon^2_2 = det [ ({\bm \alpha} +{\bm  \beta})({\bm \alpha} -{\bm  \beta})]
\\
&=&
16(A^2-B^2)(C^2-D^2),
\end{eqnarray*}
from which we deduce that 
$$\epsilon_{1/2}^2=\frac{1}{4 \#^2 } \left[C(A\pm B) \pm \frac{A^2-B^2}{2 BC} D^2 \right] .$$
Approximating $A$, $B$, $C$ and $D$ using Eq.(\ref{J1J2spinwaveH}), we find
\begin{eqnarray*}
\epsilon_{1/2}^2
&=&
\frac{J_2^2}{4 \#^2 }
\Big[2(1+\cos k+  X \cos(k/2)) (1- \cos k +X^2) 
\\
& &
-\frac{2(1+\cos k)^2}{2 X \cos(k/2) (1-\cos k + X^2)} 16 X^2 \mu^2 \sin^2 (k/2) \Big]
\\
&\approx&
J_2^2\left[ (k^2/2 +X^2)
-\frac{4 X \mu^2 k^2 }{  k^2/2 + X^2} \right].
\end{eqnarray*}
The zero point energy given by Eq.(\ref{ZeroPointEnergy}) is obtained after subtracting
$$
- \frac{1}{2 \#} (A+C)= - \frac{1}{2 \#} J_2 [ 1+\cos k +1-\cos k]=- \frac{1}{ \#} J_2 .
$$

\section{Antiferromagnet} \label{App: Antiferromagnet}
The expectation of the antiferromangentic Hamiltonian with the ansatz Eq.(\ref{SimpleMPS}) 
is given by
\begin{eqnarray*}
{\cal H}
&=&
J \sum_{i} 
 \left( \lambda_{i}+
  \lambda_{i-1} n^z_{i-1} n^z_{i+1} \right)  {\bf l}_{i}. {\bf l}_{i+1}   \\
 & &
 - J \sum_{i} \lambda_{i}
 \frac{ n_{i}^x+i n_{i}^y }{2}
  \langle {\bf l}_{i},{\bf l}_{i+1} | 
  {\bm \sigma}. {\bm \sigma} 
  |  - {\bf l}_{i},- {\bf l}_{i+1}  \rangle 
   + c.c.
\end{eqnarray*}
Its saddle point is described in the main text. An expansion about this saddle point is most easily achieved by first making a transformation to Ne\'el order parameter - changing the sign of ${\bm l}$ on alternate sites. The result (allowing for $n^y=0$ at the saddle point) is
\begin{eqnarray*}
{\cal H}
&=&
-J \sum_{i} 
 \left( \lambda_{i}+
  \lambda_{i-1} n^z_{i-1} n^z_{i+1} \right)  {\bf l}_{i}. {\bf l}_{i+1}   \\
 & &
 - J \sum_{i} \lambda_{i}
n_{i}^x
\left(
{\bm \theta}_i \cdot {\bm \theta}_{i+1} + {\bm \phi}_i \cdot {\bm \phi}_{i+1}
\right).
\end{eqnarray*}
Fixing ${\bm l}$, we find a resultant transverse-field Ising Hamiltonian for ${\bm n}$ . This is away from criticality. We focus, therefore, on the fluctuations of ${\bm l}$ about its saddle point value, keeping ${\bm n}$ fixed. Using the spin-wave expansions Eq.(\ref{SpinwaveExpansion}), we find
\begin{eqnarray*}
{\cal H}
&=&
-\frac{J}{2} \sum_{i} 
 \left( \lambda_{i}(1+n_i^x)+
  \lambda_{i-1} n^z_{i-1} n^z_{i+1} \right) 
  \left( {\bm m}_i - {\bm m}_{i+1} \right)^2.
  \end{eqnarray*}
This is identical to the usual ferromagnet with coupling constant that alternates in strength between bonds. 
Care must be taken with the Berry phase which takes the usual spin-wave form in ${\bm l}$ 
with a prefactor of ${n^z}^2$. Unlike the situation near $J_1=0$, considered in Sec. \ref{Sec:J1J2Model} and Appendix \ref{App: J1J2model}, this prefactor cannot be ignored and changes the mapping to bosonic modes. After such a mapping and diagonalization of the resulting bosonic Hamiltonian by a Bogoliubov transofrmation, we find the dispersion
$$
\omega_l
\propto
\frac{
\sqrt{
 \left[  (1 - \lambda)( 1+n^x )+\lambda {n^z}^2 \right]
  \left[  \lambda( 1+n^x ) + (1-\lambda) {n^z}^2  \right]
 }
 }
{n^z  },
$$
which reduces to the form given in the main text near the saddle point value of $n^z$.

\bibliography{bibliography}

\end{document}